\documentclass[12pt, runningheads,a4paper]{llncs}
\usepackage{amssymb}
\usepackage[colorinlistoftodos]{todonotes}
\usepackage{bbm}
\usepackage{centernot}
\usepackage{enumerate}
\usepackage{breqn}
\usepackage{soul}
\presetkeys%
{todonotes}%
{backgroundcolor=yellow}{}

\usepackage{setspace}
\usepackage{csquotes}
\usepackage[misc]{ifsym}
\usepackage{amsmath, array}
\usepackage{amssymb}
\usepackage{times}
\usepackage{xcolor}
\usepackage{bbm}
\usepackage{caption}
\usepackage{subcaption}
\usepackage{stackengine}
\usepackage{multirow}
\usepackage{pifont}
\usepackage{enumitem}
\usepackage{hyperref}
\usepackage{tabularx}
\usepackage{booktabs}
\usepackage{caption} 
\usepackage{xurl}
\usepackage{ragged2e}
\newcolumntype{F}[1]{>{\Centering}p{#1}}
\usepackage[misc]{ifsym}
\usepackage{amsmath, array}
\usepackage{amssymb}
\usepackage{times}
\usepackage{xcolor}
\usepackage{bbm}
\usepackage[a4paper, margin=1in]{geometry}
\usepackage{longfbox}
\usepackage{stackengine}
\usepackage{multirow}
\usepackage{pifont}
\usepackage{enumitem}
\usepackage{hyperref}

\begin{document}

\mainmatter  

\title{Inferring Stochastic Group Interactions within Structured Populations via Coupled Autoregression}
\titlerunning{Inferring Group Interactions within Structured Populations}

\author{\normalsize Blake McGrane-Corrigan\thanks{\textit{\small Corresponding Author} \\ (blake.mcgranecorrigan.2017@mumail.ie) \hfill}, Oliver Mason, Rafael de Andrade Moral}

\authorrunning{Inferring Group Interactions within Structured Populations}

\institute{\normalsize Department of Mathematics and Statistics,\\
Maynooth University, Kildare, Ireland}

\toctitle{Inferring Stochastic Group Interactions within Structured Populations via Coupled Autoregression}
\tocauthor{}

\maketitle

\begin{abstract}
\normalsize The internal behaviour of a population is an important feature to take account of when modelling their dynamics. In line with kin selection theory, many social species tend to cluster into distinct groups in order to enhance their overall population fitness. Temporal interactions between populations are often modelled using classical mathematical models, but these sometimes fail to delve deeper into the, often uncertain, relationships within populations. Here, we introduce a stochastic framework that aims to capture the interactions of animal groups and an auxiliary population over time. We demonstrate the model's capabilities, from a Bayesian perspective, through simulation studies and by fitting it to predator-prey count time series data. We then derive an approximation to the group correlation structure within such a population, while also taking account of the effect of the auxiliary population. We finally discuss how this approximation can lead to ecologically realistic interpretations in a predator-prey context. This approximation also serves as verification to whether the population in question satisfies our various assumptions. Our modelling approach will be useful for empiricists for monitoring groups within a conservation framework and also theoreticians wanting to quantify interactions, to study cooperation and other phenomena within social populations.

\vspace{3mm}


\end{abstract}

\section{Introduction}
Many animals have a tendency to congregate into clusters or distinct groups across time and space. This structure in turn affects the ability of such populations to compete for resources, occupy home-ranges and participate in community interactions (Oro, 2020). In the evolutionary theory of Hamilton (1964), living in groups has major benefits for group members, as well as the overall population. There are many benefits of cooperation within social populations, for example alloparenting (Lukas \& Clutton-Brock, 2018), along with many costs, such as free-riding and inter-group competition (Traulsen \& Hauert, 2009). Thus, nonlinear trade-offs exist that balance factors which either lead to higher group fitness or to the detriment of individual members. As noted by Alexander (1974), other important factors influencing group dynamics include predation risk, as well as diet and the spatiotemporal heterogeneity of resources in many large animal populations (Gittleman, 1989; Macdonald, 1983; von Schantz, 1984). Many studies have investigated the efficiency of group living (Brown, 1982; Park \& Maynard-Smith, 1990; Goodale et al., 2020; Avila \& Mullon, 2023). For example Escobedo et al. (2014) studied a mathematical model of wolf-pack hunting. They concluded that larger packs can reduce the effectiveness of group hunting, which explains why hunting success tends to peak at small pack sizes, for example. Other authors have focused on how group size may impact disease spread (Brandell et al., 2021) or theoretical aspects such as the evolution of cooperation (Nowak, 2012; Kingma et al., 2014). 

\vspace{3mm}

Our main goal is to infer changes in interactions (or associations) of animal groups within a population and how an auxiliary population (one that this population is known to associate with) may affect these interactions. This auxiliary population may be specified differently in certain contexts. For example this could be a resource, competitor, predator or prey species relative to the main population of interest. A plethora of models exists, both mathematical and statistical, for studying interactions among populations. One possibility for modelling pairwise interactions between groups may be through using Poisson Lotka-Volterra processes, originally proposed for modelling the dynamics of $n \in \mathbb{Z}_+ := \{0,1,2,...\}$ species (Schreiber, 2017). Here the mean of an inhomogenous Poisson process, which describes the changes in abundance of some population $i \in \{1,...,n\}$ is modelled by a Lotka-Volterra system. If one was to use this to model the dynamics of $n$ subgroups within a population, then the number of parameters to estimate over all the groups include $n$ intrinsic growth rates and $n^2$ interaction parameters. This does not include the number of parameters that would need to be estimated when an auxiliary population model is also considered. Therefore for inferring species interactions, this would be beneficial to explicitly deduce which groups are strongly/weakly associated to one another. A disadvantage of this approach would be that for small observational time periods, the resolution of data would be too small to estimate these interactions with reasonable certainty, with parameter estimation becoming difficult as $n$ gets large. We will propose a coupled autoregressive model for modelling group dynamics, in order to also reduce the number of parameters which need to be estimated.

\vspace{3mm}

A common feature of many count time series models is the assumption of normality on the log scale. That is, one assumes density/abundance is multiplicative on the log scale and counts are large enough so that they can be modelled by some autoregressive model (a Gompertz model for example) with some white noise error structure (Ovaskainen et al., 2017). Firstly, when counts are low, assuming normality on the log scale can be an erroneous assumption to make, as one is assuming that there is a linear relationship within the covariates and response (O’Hara \& Kotze, 2010). Secondly, taking log transformations on the raw data scale alters the interpretability of the model outputs. Lastly, one cannot account for zero observations in a straightforward manner. Typically one adds a small constant to each count to accomodate for zero observations, thus biasing the model being implemented. As noted by O’Hara \& Kotze (2010), models whereby transformations were carried out before analysis generally performed worse than Poisson and negative binomial models which had small bias also, except when there was low dispersion and mean counts were large enough to assume normality. In the context of modelling groups within a population, low counts can arise. For example, if one is studying large social mammals, groups do not usually exceed tens of individuals (Brandell et al., 2020).

\vspace{3mm}

To quantify interactions we take a stochastic approach, proposing a count model of social population dynamics. Count time series that can be classified into observation-driven and parameter-driven types were first introduced in Cox (1955). We use an observational-driven modelling approach. Given a time series $(Y_t)_{t \in [0,T]}$, for some finite time horizon $[0,T] \subset \mathbb{Z}_+$, we gather all the past information up to a time $s\geq0$ using the $\sigma$-algebra generated by observations $\{Y_1,...,Y_s\}$, given by $\cal{F}$$_s:=$ $\sigma\{Y_1,...,Y_s\}$. Given some parameter at time $t$, $\theta_t$, we assume \[\mathbb{P}(Y_t =y| \mbox{$\cal{F}$}_{t-1},\theta_t) = p(y, \theta_t) = h(y)\exp(\theta_ty-A(\theta_t)),\] where $h(y)$ some nonnegative function, i.e. $p(\cdot, \cdot)$ is an exponential family probability density function (Klenke, 2013). We also define the state process that describes the evolution of \[Z_t := \mathbb{E}(Y_t|\mbox{$\cal{F}$}_{t-1})=A'(\theta_t)\] over time, which may depend on lagged versions of $\theta_t$ and $Y_t$. $Z_t$ may be partially observed and describes the system's dynamics, inducing temporal dependencies in the data. For count time series one typically uses the canonical log link \[\ln(Z_t) = g(Z_1, ...,Z_{t-1}, Y_1, ..., Y_{t-1}),\] where $g$ is some smooth function. These models are natural extensions of general linear models that have been classically used to analyse count data in the past (Fokianos \& Tjostheim, 2011). 

\vspace{3mm}

In this paper, we will discuss the model formulation and how we can fit the model in a Bayesian setting. We then discuss different simulation scenarios that show how well our model performs. We then present an implementation of our model for inferring interactions within a real-world predator-prey system, namely wolf-elk interactions in Yellowstone National Park (YNP).We then show, under common assumptions, how we can derive a approximation to the net group correlation structure for a population, which incorporates the interactions with an auxiliary population. Finally we describe possible interpretations of this approximation in the context of predator-prey interactions.

\section{Modelling Framework}
\subsection{Coupled Autoregression}
We are interested in modelling the dynamics of $g \in \mathbb{Z}_+$ groups within some population. We also would like to know how these groups in turn interact with one another and how the presence of an additional auxiliary population could alter these interactions over time. In order to do this we will first introduce a class of state-space hierarchical models. In particular we propose an autoregressive model that couples time series of each group and an auxiliary population. 

\vspace{3mm}

Let $(X_{i, t})_{[0,T]}$ and $(Y_{t})_{[0,T]}$ respectively denote abundance (nonnegative integer-valued) time series for animal group $i \in \{1,..., g\}$ and an auxiliary population, within some finite time horizon $[0,T] \subset \mathbb{Z}_+$. $(X_{i, t})_{[0, T]}$ and $(Y_{t})_{[0, T]}$ are respectively generated by temporal point processes $X_i$ and $Y$. Let $\cal{F}$$_{t}$ denote the internal histories of these processes up to time $t$ and \[\mbox{$\cal{H}$$_{t}$} := \sigma \{X_{1,t}, ..., X_{g,t}, Y_t\}\] denote the $\sigma$-algebra generated by $X_{i,t}$ and $Y_t$ at time $t$, for $i \in \{1,...,g\}$. We first assume that for all $i \in \{1,...,g\}$ and any $t \in [1,T]$ \begin{eqnarray*}&&\mathbb{P}( X_{i,t} | \mbox{$\cal{F}$$_{t-1}$} ) = \mathbb{P}( X_{i,t} | \mbox{$\cal{H}$$_{t-1}$}) \\ &&\mathbb{P}( Y_{t} | \mbox{$\cal{F}$$_{t-1}$}) = \mathbb{P}( Y_t | \mbox{$\cal{H}$$_{t-1}$}),\end{eqnarray*} i.e. $\left(X_{i,t}\right)_{[0,T]}$ and $\left(Y_{t}\right)_{[0,T]}$ are first-order Markovian. We then model $(X_{i, t})_{[0,T]}$ and $(Y_{t})_{[0,T]}$ as conditional first order inhomogenous Poisson ($\cal{P}$) processes. That is for each $t \in [1,T]$ and $i \in \{1,...,g\}$
     \begin{eqnarray} \label{model2}
	    &&X_{i,t} | w, \mbox{$\cal{H}$}_{t-1} \sim \mbox{$\cal{P}$}\left(\mu_{i,t}^{X} \right) \\ 
     \label{model1}
        &&Y_{t} | w, \mbox{$\cal{H}$}_{t-1} \sim \mbox{$\cal{P}$}\left( \mu_{t}^{Y} \right).
    \end{eqnarray}
    \noindent $X_{i,0}=x_{i,0}$ and $Y_{0}=y_0$ are assumed to be known a priori. In (\ref{model2}) and  (\ref{model1}) we have that \begin{equation} \label{w}
    \begin{aligned}
    &w:=\{\omega_i^X, \lambda_i^X, \psi_i, \delta_i; 1 \leq i \leq g\}
    \end{aligned}
\end{equation} are mutually independent random effects, where
     \begin{equation} \label{re}
    \begin{aligned}
    &\omega_{i}^{X} \sim \mbox{$\cal{N}$}(\mu_{\omega}, \sigma^2_{\omega}), \ \ \lambda_{i}^{X} \sim \mbox{$\cal{N}$}(\mu_{\lambda}, \sigma^2_{\lambda}), \\
    &\psi_{i} \sim \mbox{$\cal{N}$}(\mu_{\psi}, \sigma^2_{\psi}), \ \  \delta_{i} \sim \mbox{$\cal{N}$}(\mu_{\delta}, \sigma^2_{\delta}),
    \end{aligned}
\end{equation} for each $i \in \{1,...,g\}$. We therefore have that \begin{equation} \label{meanvar}
    \begin{aligned}
    &M:= \{\mu_{\omega}, \mu_{\lambda}, \mu_{\psi}, \mu_{\delta}\} \subset \mathbb{R}^4 \\ 
    &\Sigma := \left\{\sigma^2_{\omega}, \sigma^2_{\lambda}, \sigma^2_{\psi}, \sigma^2_{\delta}\right\} \subset \mathbb{R}_+^4
\end{aligned}
\end{equation} are respectively the set of population level means and variances for (\ref{re}). These random effects are assumed to be mutually independent from another and to the elements of $\mbox{$\cal{H}$}_{t-1}$. These will be included in the components that make up the linear predictors for $\mu_{i,t}^X$ and $\mu^Y$. The inclusion of random effects will allow us to account for across-group variability in each of the mean processes. The log-intensity process for (\ref{model2}) is given by the recursion \begin{align} \label{loglink2}
        &\ln(\mu_{i,t}^{X}) = \omega^{X}_{i} + \Lambda^{X}_{i,t-1} + \Psi_{i,t-1} + \Delta_{i,t-1},
    \end{align} where $\omega^{X}_{i}$ is the random intercept parameter for group $i$, the autoregressive component 
     \begin{equation*}
     \begin{aligned}
        &\Lambda^{X}_{i,t} = \lambda_i^{X} \ln\left(x_{t} + 1\right), \end{aligned}
    \end{equation*} accommodates temporal correlation in $X_{i,t}$, the intraspecific component \begin{equation} \label{psi}
    \begin{aligned}
        &\Psi_{i,t}  = \psi_i \ln\left( \sum_{j \not = i}^g x_{j,t} + 1\right),
    \end{aligned}
    \end{equation} accounts for how cumulative changes in abundances of each group $j \neq i$ affects group $i$, and the auxiliary component \begin{equation} \label{delta}
    \begin{aligned}
        &\Delta_{i,t} = \delta_i \ln\left(y_{t} +1\right),
    \end{aligned}
    \end{equation} accounts for how changes in the auxiliary population affect group $i$. The log-intensity process for (\ref{model1}) is given by the recursion \begin{align} \label{loglink1}
        &\ln(\mu_{t}^{Y}) = \omega^{Y} + \Lambda^{Y}_{t-1} + \Gamma_{t-1},
    \end{align} where $\omega^{Y} \in \mathbb{R}$ is an intercept parameter, the autoregressive component \begin{equation*}
    \begin{aligned}
        &\lambda^{Y}_{t} =  \lambda^{Y} \ln\left(y_{t} + 1\right), \end{aligned}
    \end{equation*} accommodates temporal correlation in $Y_t$, with $ \lambda^{Y}  \in \mathbb{R}$, and the interspecific component \begin{equation} \label{gamma}
    \begin{aligned}
        &\Gamma_{t}  = \gamma \ln\left(\sum_{j=1}^g x_{j,t} +1\right),
    \end{aligned}
    \end{equation} accounts for how changes in the total group population affects $Y_t$, where $\gamma \in \mathbb{R}$. Note that the inclusion of a log-link in (\ref{loglink2}) and (\ref{loglink1}) is to ensure that $\mu_{i,t}^{X}$ and $\mu_t^Y$ are non-negative. Adding a constant $1$ within each $\ln$ argument ensures that our model can capture zero abundance observations and so correlations can take values in $\mathbb{R}$, as noted by Fokianos and Tjostheim (2011).

    \vspace{3mm}

    Conditioning on (\ref{re}) in the latent processes (\ref{loglink2}) and (\ref{loglink1}) allows for the accommodation of overdispersion, i.e. violation of expectation-variance equality (Molenberghs, 2010), a common issue faced when modelling count time series. Note that our modelling approach is similar to level-correlated models, which incorporate dependence between components of a count vector via an underlying correlated latent process (Aitchison and Ho, 1989; Chib and Winklemann, 2001; Davis et al., 2021).  Another way of allowing for overdispersion is using quasi-likelihood methods, but we adopt a mixed model approach in order to also account for the grouping behaviour and inter-group variation. In time series models of counts one routinely includes observational and latent processes to account for measurement error (Hosack, Peters and Hayes, 2012). In our hierarchical framework, the data generating process is given by (\ref{model2}) and (\ref{model1}), with the latency modelled by (\ref{loglink2}) and (\ref{loglink1}). We do not explicitly model the measurement process, as we are only concerned with inferential aspects of our model.

\subsection{Bayesian Inference}

To fit our model we took a Bayesian perspective. In a genereal setting one is typically interested in calculating the posterior density for a parameter vector, $\theta$, given an observation vector, $y$. If this cannot be analytically computed, one can specify a prior distribution, $\pi(\theta)$. This is then updated via some Monte Carlo algorithm, into an approximated posterior distribution via \[p(\theta | y) = \frac{\pi(\theta) p(y | \theta)}{p(y)},\] where $p(y | \theta)$ is the likelihood and $p(y)$ is the marginal distribution of $y$ (Klenke, 2013). In all of the analyses below, we employed a Hamiltonian, no-U- turn sampler (NUTS) Markov chain Monte Carlo (MCMC) algorithm to fit our model. We implemented this via the probabilistic programming language Stan, using the R package rstan (Carpenter et al., 2017). We used four chains in each simulation of our simulation scenarios and case study below. For our case study we used 20,000 iterations, a thinning rate of 20 and burning of 2000. Convergence was assessed using trace plots, autocorrelation function plots, effective sample size and the Gelman-Rubin statistic (Gelman and Rubin, 1992). 

\vspace{3mm}

We assigned independent noninformative Gaussian priors to our mean hyperparameters and auxiliary population parameters, \[\theta_{\mu} \sim \mbox{$\cal{N}$}(0,100^2), \ \theta_{\mu} \in M \cup \left\{\omega^Y, \lambda^Y, \gamma \right\},\] where $M$ is given in (\ref{meanvar}). According to Gelman (2006) if one was to set a variance prior such as $\mbox{Inv-}\Gamma(\epsilon, \epsilon)$, for $\epsilon$ small, then there still does not exist a proper (integrates to unity) limiting posterior distribution. Thus this distribution is quite sensitive to changes in values of $\epsilon$. For noninformative proper priors Gelman (2006) recommend using a half-Gaussian distribution, $\mbox{$\cal{N}$}^+(0, \sigma^2)$ for large $\sigma^2$. Thus for our standard deviation hyperparameters we assigned independent half-Gaussian priors, \[\theta_{\sigma} \sim \mbox{$\cal{N}$}^+(0,100^2), \ \theta_{\sigma} \in \Sigma,\] where $\Sigma$ is given in (\ref{meanvar}). Note that if one wanted to give more weight to the tails, one could use truncated t-distributions as priors for variance parameters. The inclusion of heavier-tailed distributions as proper priors may result in one's model being robust to outlying observations (Wakefield et al., 1994), but as noted by O'Hagan and Pericchi (2012) the utility of heavy-tailed models in complex settings is not so clear.

\begin{table}[h]
\centering
\begin{tabular}{ccccccccc}
    \hline
    {\textbf{Parameter}} \ \  & $\omega^Y$ \ \ & $\lambda^Y$ \ \ & $\mu_{\omega}$ \ \ & $\mu_{\lambda}$ \ \ & $\sigma_{\omega}$ \ \ & $\sigma_{\lambda}$ \ \ & $\sigma_{\delta}$ \ \ & $\sigma_{\psi}$\\
    \hline
    {\textbf{True Value}} \ \ & 0.9 \ \ & 0.9 \ \ & 0.9 \ \ & 0.06 \ \ & 0.2 \ \ & 0.2 \ \ & 0.04 \ \ & 0.04\\
    \hline
\end{tabular}
\caption{True parameter values chosen for simulation scenarios S1-S5.}
\label{params}
\end{table}

\section{Simulation Study}

To evaluate model performance we simulated data from (\ref{model2}) and (\ref{model1}) for 5 different scenarios:
\begin{enumerate}
    \item $g=10$, $T=20$;
    \item $g=42$, $T=20$;
    \item $g=50$, $T=50$;
    \item $g=42$, $T=20$, $\gamma = -0.15$, $\mu_{\psi}= 0.15$, $\mu_{\delta} = 0.15$
    \item $g=42$, $T=20$, $\gamma = 0.05$, $\mu_{\psi}= -0.05$, $\mu_{\delta} = -0.05$.
\end{enumerate}
Scenarios 1 - 3 were used to observe how well parameters were estimated when the resolution of our data varies from low  to high. We included the last two scenarios to deduce how well parameters are estimated where one is respectively monitoring, say, a predator population and its prey. In scenario 4 it is assumed that the predator has a strong positive net effect of changes in prey abundance on predator groups, strong negative net effect of changes in predator abundance on prey and weak net effect of changes in predator abundance on groups. In scenario 5 it is assumed that the predator has a weak positive net effect of changes in prey abundance on groups, weak negative net effect of changes in predator abundance on prey and strong net effect of changes in predator abundance on groups. 

\vspace{3mm}

We ran 100 simulations for each scenario, with parameters set as in Table \ref{params}. For scenarios 1-3 we set $\gamma=-0.05$, $\mu_{\psi} = -0.5$ and $\mu_{\delta} = 0.2$. For scenarios $1,2,4$ and $5$ we used 20,000 iterations, a thinning rate of 20, burnin of 2000. For the large scenario we used 40,000 iterations, a thinning rate of 40 and burnin of 4000. The accuracy of parameter estimates was assessed by calculating the root mean square error (RMSE), bias (B) and relative bias (RB) over all $100$ simulations. Let $\theta$ be a true (known) parameter and $\hat{\theta}_i$ an estimate for simulation $i \in \{1,...,n\}$. We define RMSE over $n$ simulations as
\[\mbox{RMSE} := \sqrt{\frac{\sum_{j=1}^n (\hat{\theta}_j-\theta)^2}{n}}.\] We define B and RB for the $i$th simulation as \begin{eqnarray*}
    &&\mbox{B}_i :=\hat{\theta}_i-\theta \\ 
    &&\mbox{RB}_i := \frac{B_i}{\theta}.
\end{eqnarray*}

RMSE assesses how close our predicted values are to the true values. The lower RMSE is the closer our estimates are to the true simulated parameters. One issue with RMSE is that it is scale-dependent. Therefore we also used B and RB as other accuracy measures. B and RB, in our context, measures the difference and scaled difference between the mean obtained from a large number of simulations and our true parameter value. Unlike RMSE, RB is not sensitive to anomalously large/small values that may arise in simulation studies. Thus RB can be robust and a better accuracy metric in a simulation context. For all scenarios the RMSE for parameter estimates can be seen in Table \ref{rmse1}. We clearly see an improvement of parameter estimation, via lower values of RMSE, as our sample size increases from scenarios 1 to 3, with some possible small exceptions. The RMSE for $\mu_{\omega}$ gets smaller as sample size increases but remains relatively large at 0.40627. For scenarios 4 and 5 we see that in general RMSE is quite low, albeit for $\mu_{\omega}$ in scenario 4. This large value could be due to outlier values as predicted by our model in certain simulation runs.

\begin{table}[h]
\centering
\begin{tabular}{cccccc}
   \hline
    \multirow{2}{1.8cm}{\centering {\textbf{Parameter}}} \ \ & \multicolumn{5}{c}{{\textbf{RMSE}}} \ \\
    \cline{2-6}
    & {\textbf{1}} \ \ & {\textbf{2}} \ \ & {\textbf{3}} \ \ & {\textbf{4}} \ \ & {\textbf{5}} \\
    \hline
    $\omega^Y$ & 0.061 & 0.057 & 0.059 & 0.053 & 0.272  \\
    $\lambda^Y$ & 0.009 & 0.008  & 0.007  &  0.004 & 0.015  \\
    $\gamma$ & 0.011 & 0.008 & 0.006  &  0.005 & 0.025  \\
    $\mu_{\omega}$ & 0.784 & 0.472 & 0.406  &  1.187 & 0.135  \\
    $\mu_{\lambda}$ & 0.129 & 0.064 & 0.078 & 0.068 & 0.042 \\
    $\mu_{\psi}$ & 0.120 & 0.079 &  0.102 & 0.099  & 0.028 \\
    $\mu_{\delta}$ & 0.088 & 0.055 & 0.027 & 0.091 & 0.007 \\
    $\sigma_{\omega}$ & 0.147 & 0.060 & 0.069 &  0.089  & 0.060 \\
    $\sigma_{\lambda}$ & 0.109 & 0.066 & 0.044 & 0.069 & 0.031  \\
    $\sigma_{\delta}$ & 0.011 & 0.012 & 0.014 & 0.014 & 0.007 \\
    $\sigma_{\psi}$ & 0.015 & 0.015 & 0.019 & 0.024  & 0.012 \\
    \hline
\end{tabular}
\caption{RMSE for estimates in scenarios 1 - 5 calculated across 100 different simulations.}
\label{rmse1}
\end{table}

For B and RB we plotted boxplots, with medians and outliers, as can be seen in Figures \ref{small_box} - \ref{comphigh} in the Appendix. In all the scenarios we can see that the boxplots for B and RB have medians close to $0$. Issues arise however with the biases for the intercept mean in scenario 5, as data resolution increases. We therefore plotted bivariate plots of our MCMC draw in order to see if there was any confounding between parameters (see Figures \ref{biv1} - \ref{biv3} in the Appendix for a single simulation for S1-S3 as an example of such confounding). In scenarios scenarios 1-3 we saw slight confounding between $\gamma$ and $\lambda^Y$, and between $\mu_{\delta}$ and $\mu_{\omega}$, which indicates that these parameters are slightly biased and there may be small identifiability issues, as already indicated through the larger RMSE values for $\mu_{\omega}$. This confounding reduced as the sample size increased, showing that larger resolution of data results in less biased estimation (see Figures \ref{biv1} - \ref{biv3}). RB are consistently small, indicating that under all of these scenarios of interest our model is able to accurately estimate parameter values with little bias. Note that there were relatively small biases for the standard deviation parameters in particular. Therefore, together with the overall low RMSE values, small RB values indicates that our estimates are close enough to their true values, with relatively small bias.

\section{Predator-Prey Case Study}

We will now apply our model to a case study, namely the predator-prey dynamics of grey wolves (\textit{Canis lupus}) and elk (\textit{Cervus canadensis}) in Yellowstone National Park (YNP) over 20 years. Since the reintroduction of grey wolves to YNP and the surrounding areas in 1995, wolf and elk populations have been thoroughly monitored in order to assess ecosystem responses (Boyce, 2018). Thus a lot is known about their predator-prey relationship and, more generally, the ecological processes that affect these populations. Predation is one of several regulatory factors of elk population growth in YNP (Smith, Stahler and MacNulty, 2021). By applying predator-prey theory, Varley and Brewster (1992) predicted that elk abundance would decrease to a low equilibrium state following wolf reintroduction. Many studies have questioned how significant this predation effect has been (Peterson et al., 2014). However there is general consensus that wolf predation is largely density-dependent (Vucetich, Peterson \& Schaefer, 2002), with stronger dependence attributed wolf density. 

\begin{figure}[!h]
\centering
\includegraphics[clip, width = 14cm]{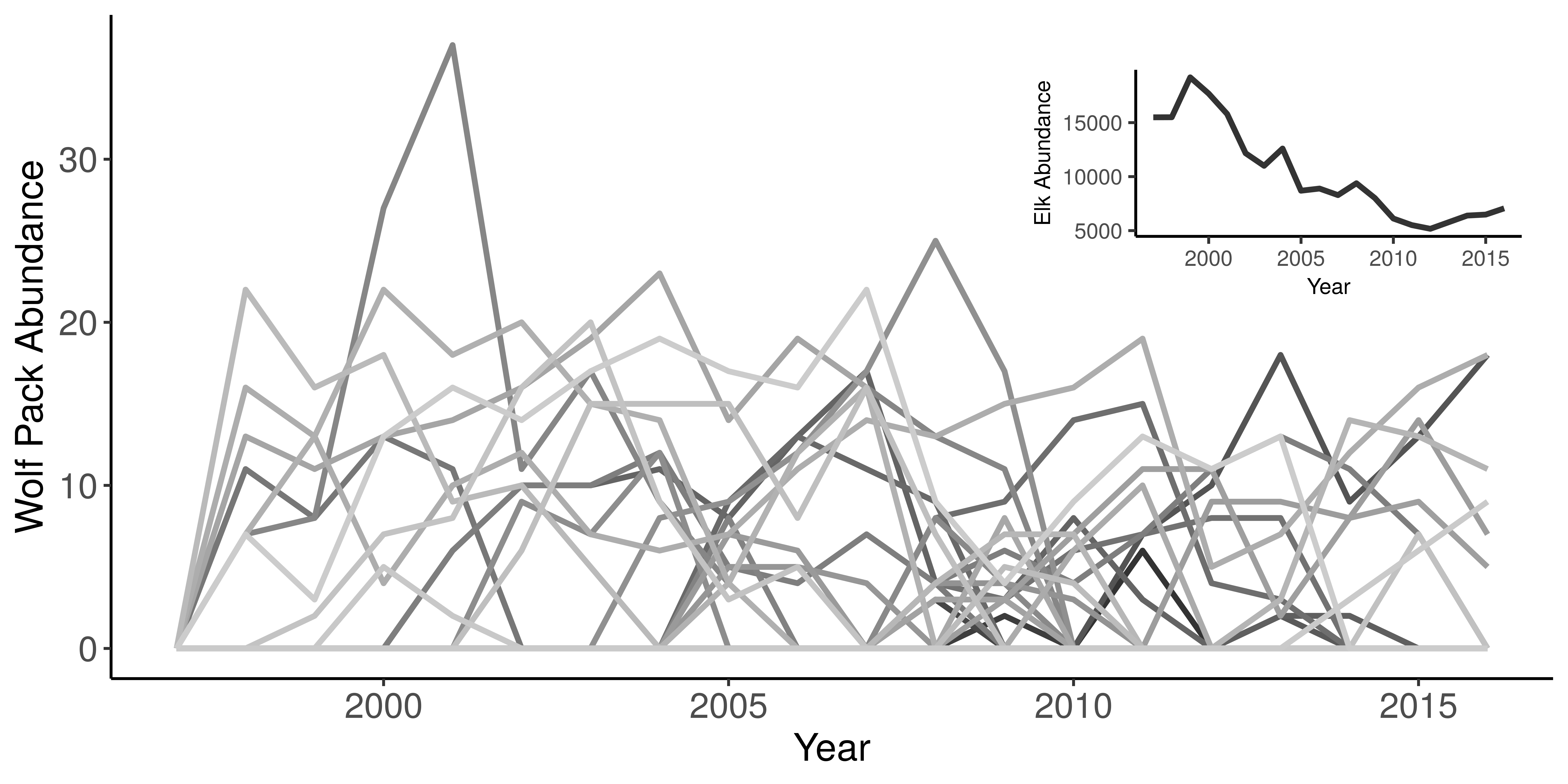}
\caption{Time series of wolf packs (main) and elk (inset) abundances in YNP from 1996-2016.}
\label{wolfelkplot}
\end{figure}

It has been observed that the rate of predation of wolf packs can remain constant even when elk densities are declining, suggesting that at certain times the functional response of wolves may just be a linear function of prey abundance (Hebblewhite et al., 2018). Intra-pack and inter-pack competition also has a substantial effect on hunting efficiency and prey acquisition by wolf packs (Cassidy et al. 2015). This suggests that studying pack dynamics in this context is vital to understand how social animals behave over time. 

\vspace{3mm}

Annual predator and prey observational abundance data was obtained from Brandell et al. (2021) and altered for time series analyses (see Fig. \ref{wolfelkplot}). Yearly abundance data for grey wolf packs from YNP were collated over 20 years using observational studies and field research. As noted by Brandell et al. (2021), in the dataset, the membership and composition of packs is known. For our framework, we therefore assume that we have perfect detection, i.e. we do not have an additional process that explicitly takes account of the uncertainty of abundance measurement. If one's goal was to estimate wolf abundance over time, one may use temporal N-mixture models, that allow for detection error and measurement uncertainty (Mimnagh et al., 2022).

\subsection{Group Formation and Splitting}
Grey wolf packs are known to form and disband over time via natal or breeding dispersal (Smith, Stahler and MacNulty, 2021; Morales-Gonzalez et al., 2022). They are largely territorial with overlapping spatial ranges. However, wolf packs tend to avoid neighbouring packs, in order to miminise aggressive interactions and mortality (Kauffman et al., 2007). We assume that once a pack forms, its abundance stays positive until it splits, and thereafter its abundance is zero. In order to account for pack formation and splitting we modify (\ref{loglink2}) to become
\begin{eqnarray*}
    &&X_{i,t} | w, \mbox{$\cal{H}$}_{t-1} \sim \mbox{$\cal{P}$}\left(\mu_{i,t}^XC(X_{i,t})\right) \\
    &&C(X_{i,t}) :=  \underbrace{\mathbbm{1}\{ X_{i,t-1}>0\}}_{A(X_{i,t})} + \underbrace{\mathbbm{1}\left\{ \sum^{t-1}_{k=1}X_{i,k}=0\right\}}_{B(X_{i,t})},
\end{eqnarray*} where $\mu_{i,t}^X$ is the same as in (\ref{loglink2}). Here $\mathbbm{1}\{\cdot\}$ is an indicator function, that is $\mathbbm{1}\{Z\} = 1$ if $Z$ holds and $0$ otherwise. To initialise our model we define $A(X_{i,0}) := 0$ and $B(X_{i,0}) := 1$, or $A(X_{i,0}) := 1$ and $B(X_{i,0}) := 0$, so that $C(X_{i,0}) := 1$. In the wolf-elk data-set obtained from Brandell et al. (2021), observations of lone wolves (members of no packs) and transient packs (groups existing for less than two consecutive months) were removed by the authors. Therefore we do not want to model (non-group) events such as these. 

\begin{table}[h]
\centering
\begin{tabular}{ccccc}
    \hline
    $\mathbf{t}$ \ \ & $\mathbf{X_t}$ \ \ & $\mathbf{A(X_t)}$ \ \ & $\mathbf{B(X_t)}$ \ \ & $\mathbf{C(X_t)}$ \ \ \\
    \hline
    $0$ & $0$ & $0$  & $1$ & $1$\\
    $1$ & $0$ & $0$ & $1$ & $1$\\
    $2$ & $3$ & $0$ & $1$ & $1$ \\
    $3$ & $10$ & $1$ & $0$ & $1$ \\
    $4$ & $5$ & $1$ & $0$ & $1$ \\
    $5$ & $0$ & $1$ & $0$ & $1$ \\
    $0$ & $0$ & $0$ & $0$ & $0$\\
    $0$ & $0$ & $0$ & $0$ & $0$\\
    \hline
\end{tabular}
\caption{Example of how $A$ and $B$ allow for group formation and splitting, where $X_t$ is a count random variable at time $t\geq0$.}
\label{ind}
\end{table}

Our model is first order Markov and so at each time point, if a pack has not formed (pack abundance is 0 prior to this time) then there is still opportunity for pack emergence. We thus want to include these zero abundance observations in our extended framework. Analogously, if a pack has split (pack abundance is zero after this time) then we know it will not form again. Thus we want to ignore these zero abundance observations. We include an illustrative example of how the indicators $A$ and $B$ respectively correspond to pack formation and splitting in Table \ref{ind}. If one was interested in explicitly modelling how dispersers move between packs, one could potentially model this with spatial birth-death-movement processes (Lavancier \& Le Guével, 2021) or process convolution models within a hierarchical framework (Hooten et al., 2018).

\subsection{Results}
Parameters estimates can be seen in Table \ref{casetable}, along with their respective $95\%$ credible interval (CI). We can interpret our parameter estimates in the context of this wolf-elk system as characterising how each group/population is affected by the other over the observation period. Using Bayesian models for inferential purposes means one must test for parameter identifiability. Weakly identifiable models are subsets of parameter redundant models, and can therefore aid one in knowing when one's model is unidentifiable (Cole, 2020). It is important to know whether parameters are identifiable when the focus is interpreting parameter estimates, for example to infer interaction strengths. This is because we would like to know if these accurately capture the behaviour of the system in question. 

\begin{table}[h]
\centering
\begin{tabular}{lllll}
    \hline
    \textbf{Parameter}& \textbf{Estimate} & $\mathbf{95\%}$ \textbf{CI} & \textbf{PPO (\%)} & \textbf{Interpretation} \\
    \hline
    $\omega^Y$ & $0.85$ & $[0.74, 0.96]$ & $0.20$ & Moderate elk growth rate \\
    $\lambda^Y$ & $0.91$ & $[0.90, 0.93]$ & $0.00$ & Weak net intraspecific \\
    &  & & & effect of changes in elk abundance \\
    $\gamma$ & $-0.02$ & $[-0.03,-0.02]$ & $0.00$ & Weak net effect of changes\\
    &  & & & in wolf abundance \\
    $\mu_{\omega}$ & $10.32$ & $[4.05, 17.70]$ & $7.50$ & High wolf growth rate \\
    $\mu_{\lambda}$ & $0.50$ & $[0.29, 0.69]$ & $0.40$ & Moderate net intraspecific\\
    &  & & & effect of changes in pack abundance \\
    $\mu_{\psi}$ & $1.14$ & $[0.56, 1.84]$ & $0.90$ & Weak inter-pack net effect\\
    &  & & & for changes in pack abundance \\
    $\mu_{\delta}$ & $-1.74$ & $[-2.55,-1.06]$ & $1.20$& Weak net effect of changes in elk \\
    &  & & & abundance\\
    $\sigma_{\omega}$ & $15.24$ & $[9.93, 21.92]$ & $6.90$ & High pack growth variability \\
    $\sigma_{\lambda}$ & $0.46$ & $[0.28, 0.7]$ & $0.30$ & Weak intra-pack variability for net\\
    &  & & & effect of  changes in pack abundance\\
    $\sigma_{\delta}$ & $1.80$ & $[0.83, 2.19]$ & $1.10$ & Moderate wolf variability for net\\
    &  & & & effect of  changes in elk abundance\\
    $\sigma_{\psi}$ & $1.41$ & $[1.20, 2.58]$ & $1.10$ & Moderate wolf variability for net \\
    &  & & & effect of changes in pack abundance\\
    \hline
\end{tabular}
\caption{Parameter estimates, with respective $95 \%$ credible interval (CI), prior-posterior overlap (PPO) and parameter interpretation for YNP wolf-elk case study.}
\label{casetable}
\end{table}

Suppose $f(\theta | y)$ is the marginal posterior distribution for a parameter, $\theta$, and $\pi(\theta)$ the prior distribution assigned to $\theta$. Then $\theta$ is weakly identifiable if \[f(\theta | y) \approx \pi(\theta).\] By plotting the prior-posterior overlap (PPO) percentage, or by computing this overlap via kernel density estimation, one can deduce whether their model is weakly identifiable (Cole, 2020). For this case study we looked at the prior-posterior overlap for each of our estimates. These can be seen in Table \ref{casetable} (graphically in Fig. \ref{identplot} of the Appendix). The largest observed overlap was $7.5\% < \tau$, where $\tau = 35\%$ is the recommended (ad hoc) threshold value of Garrett and Zeger (2000). If one has that the prior-posterior overlap for each parameter is above this threshold, then that parameter is weakly identifiable, i.e. the data does not provide us with more information than the priors alone, and thus the model is unidentifiable. As all of our parameters have sufficiently small prior-posterior overlap, we can conclude that our model is identifiable, giving us stronger justification for interpreting our parameter estimates.

\vspace{3mm}

In YNP, elk undergo habitat selection so as to avoid wolf predation in the summer. In winter, they use other strategies, like grouping, to avoid predation, thus altering their seasonal distribution in direct response to predator presence (Mao et al., 2005). For elk, the estimated intercept, $\hat\omega^Y$, and autoregressive parameter, $\hat\lambda^Y$, suggest that elk intrinsic growth was moderate and the net intraspecific effect of changes in elk abundance was weak over the study period, i.e. weak density-dependence. Many mammals, such as ungulates, are known to exhibit irruptive dynamics, i.e. short term increases in abundance followed by rapid decline (Gross, Gordon \& Owen‐Smith, 2010; Duncan et al., 2020). This is in line with our findings, suggesting that elk mean growth was high with little density dependence hindering this population from increasing in abundance. Perhaps this weak density dependence played a more significant role as the population reached higher values, when also coupled with other factors, thus reducing the overall population to a much lower level near the end of our study period. 

\begin{figure}[!h]
\centering
\includegraphics[clip, width = 15.5cm]{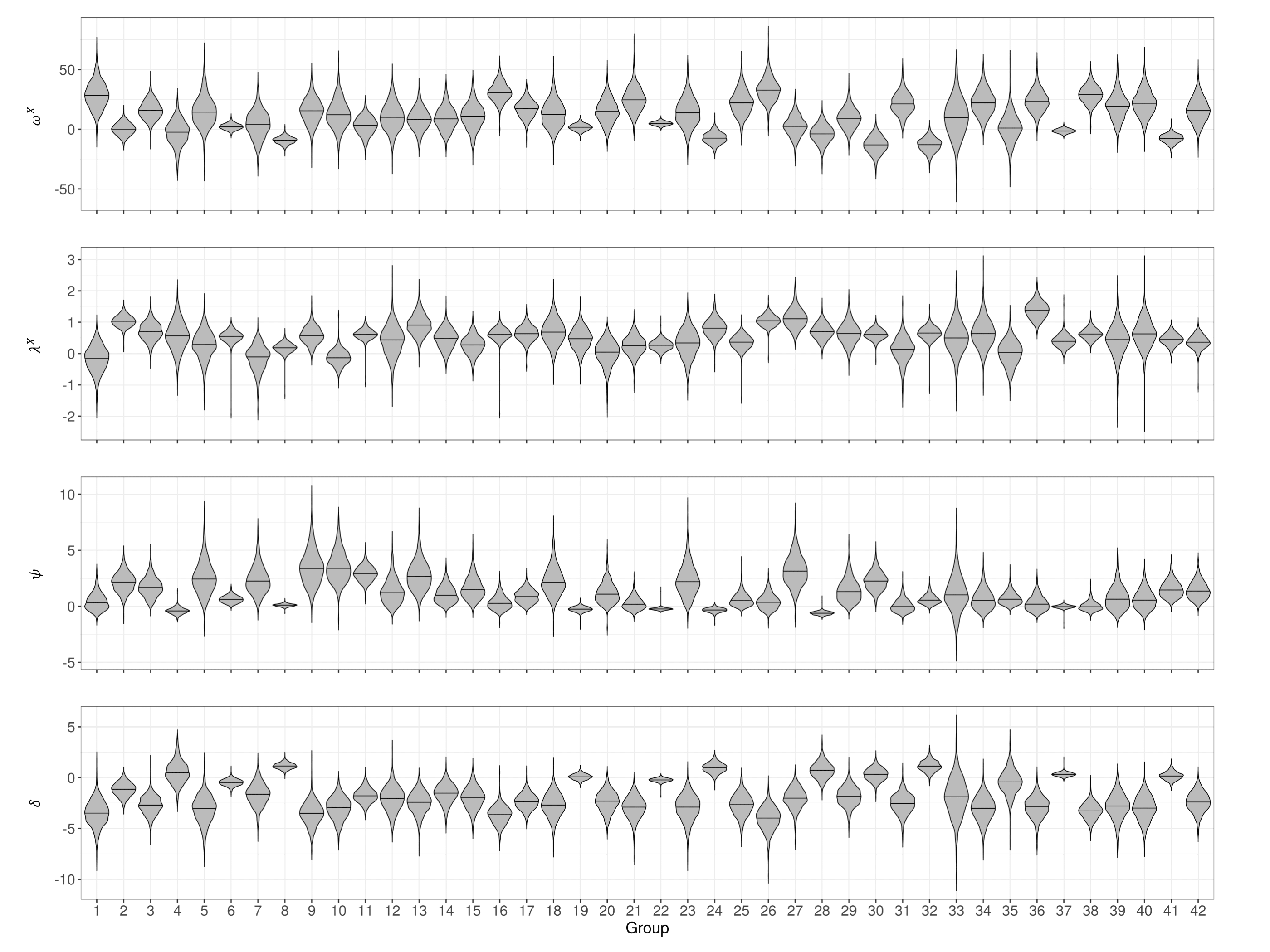}
\caption{Violin plots for $\omega^X := \{\omega^X_{1}, ..., \omega^X_{{42}}\}$, $\lambda^X := \{\lambda^X_{1}, ..., \lambda^X_{{42}}\}$, $\psi := \{\psi_{1}, ..., \psi_{42}\}$ and $\delta := \{\delta_{1}, ..., \delta_{42}\}$. Each plot shows the posterior density estimates for realisations of the random effects for all $42$ wolf packs, with the respective $0.5$-quantiles.}
\label{combplot}
\end{figure}

We interpret $\hat\gamma$ as the net effect of changes in wolf abundance on elk. Some authors have suggested that the influence of predation pressure on declining elk population has been overemphasised, with hunting pressure and drought, for example, having a substantially larger effect (Middleton et al., 2013; Peterson et al., 2014). Our results suggest that $\hat\gamma$ is small and negative, indicating that increases in the wolf population abundances has had a small net effect on the elks mean process. This suggests that over this 20 year period wolf related processes, such as predation, may not have been the dominant cause of elk abundance reduction. 

\vspace{3mm}

We can see that $\hat\mu_{\omega}$ and $\hat\sigma_{\omega}$ are both large, thus indicating that there was a high wolf population mean growth rate, with large variability, over the study period. This suggests that wolf pack abundance has increased significantly since reintroduction, with increases varying in magnitude across packs. Pack size is known to be a major determinant of hunting success, with this success varying with prey type (Smith, Stahler and MacNulty, 2021). Small wolf pack sizes result in low prey consumption, but also low kill rates, meanwhile large pack sizes result in both high kill rates and consumption (Wilmers et al., 2003). Pack composition is also an important aspect, due to quantitative benefits such as hunting success and negative consequences when acquiring resources, such as free-riding. For example, packs with a larger proportion of older members or adult males have higher odds of winning in aggressive interactions, even though they may have a quantitative disadvantage (Cassidy et al., 2015).

\vspace{3mm}

The low value of $\hat\mu_{\psi}$ suggests that there was low inter-pack net effects in response to changes in the abundance of each pack, with moderate variation around this population mean as seen via $\hat\sigma_{\psi}$. Even though wolf packs are highly territorial, our model might not capture this fully as we are not taking into account that some packs may rarely come into contact with one another, due to low overlapping ranges. It has been noted however that the specific location of the aggressive interaction in relation to pack territories has little to no effect on inter-pack interaction outcomes. Thus the relationship of aggressive interactions to spatial location is quite heterogeneous over time. The interactions of two packs may benefit each of them, perhaps via indirect mutualism (Vandermeer, 1980), at specific times depending on pack composition (Cassidy et al., 2015). 

\vspace{3mm}

\noindent The low value of $\hat\mu_{\delta}$ has multiple interpretations. As a first interpretation it may suggest that there was weak elk predation, with moderate variation around this population mean as seen via $\hat\sigma_{\delta}$. Grey wolves are generalists and so have multiple prey types at any given time. They can also undergo prey switching when certain prey types become scarce (Prokopenko, 2022). As optimal foraging theory suggests, wolves have a preference for elk, but this decreases as total prey abundance available declines. Prey selection by wolves has been shown to be primarily determined by the within-population vulnerability of elk, where wolves tend to minimise risk of harm from predation (Hoy et al., 2021). Grey wolves can also shift from hunting to scavenging when their primary prey become rare (Tallian, 2017). Secondly, the percentage of elk hunted by wolf packs has remained around 5\% per year despite elk numbers decreasing significantly over the 20 year study period (Smith, Stahler and MacNulty, 2021). This observation and the fact that total wolf abundance rapidly increased up to 2007 and then fell slightly, saturating at around 100 wolves, could also explain the negative mean estimate. The second intepretation is that this parameter may not capture direct predation. Increases in the elk population, at certain times, could have coincided with decreases in the wolf population, thus suggesting that the effect of increasing elk is negative. As in classical predator-prey theory, there are times when predator abundance will be decreasing as prey abundance increases, as there is a delay effect on their dynamics. After 2007, when the overall wolf population was around 170 individuals, abundance decreased significiantly and flucuated close to 100 individuals up to 2016 (Smith, Stahler and MacNulty, 2021). Following a rapid increase in abundance from 1997, this lulling period may explain why this parameter is low and negative.

\vspace{3mm}

We can observe the realisations of the random effects for each pack through violin plots (showing the estimated posterior density with $0.5$-quantiles) in Figure \ref{combplot}. Some packs, like pack $6$ and $19$ for example, have medians close to $0$ for both  $\mu_{\psi}$ and $\mu_{\delta}$. Meanwhile others, such as pack $23$ and $27$, have medians below $0$ for $\mu_{\delta}$ and above $0$ for $\mu_{\psi}$. Packs $4$ and $28$, for example, have random effect medians for $\mu_{\psi}$ below $0$, indicating negative net effects in relation to other packs present in the system. The resolution of time series data was different for each pack, resulting in varying levels on uncertainty around these median values, as seen in the probability density plots of our posterior estimates in Figure \ref{combplot}. Even though the population level means for each random effect may have been positive or negative, these effects varied across all $42$ packs, showing that our parameters are able to capture inter-pack heterogeneity when explicitly modelling group dynamics. The relationship between packs cannot be reduced to single time-invariant parameter values like in our model. However, our mean estimates do capture the general relationship between wolf packs and elk, with the changes in both elk and wolf abundance having quite nonlinear net effects on each pack's dynamics.

\vspace{3mm}

This predator-prey system, although highly complex, can offer us insight into how species may interact with one another, while also explicitly modelling the internal structure of one of the species. Garrott et al. (2008) concisely noted that ``as the elk population declined wolves killed a larger proportion of the population. Such an inverse density-dependent response is destabilizing, however, predator-prey theory suggests that at low prey densities the total response may become density dependent and thus be regulatory, resulting in persistence of the predator–prey system at a lower prey density than realized in the absence of wolves." These findings could explain how some packs have negative net effects to increases in elk abundance, for example. This system at times may exhibit cyclic behaviour, a common feature exhibited by the basic Lotka-Volterra predator-prey model. Time-invariant parameters may not capture such inter-pack temporal heterogeneity. Therefore in order to account for potential features such as cyclic behaviour, one could extend our model by using time-varying parameters.

\section{Quantifying Group Interactions}
\subsection{Correlation Approximation}
Our modelling framework includes (\ref{psi}), (\ref{delta}) and (\ref{gamma}) that respectively capture the effect of changes in abundance of each group with every other group, of each group with an auxiliary population and of an auxiliary population with the main population of interest. Definitions of interactions vary within the literature (Berlow, 2004). Our aim was to formulate a model that could be used to quantify changes in associations. Our model allows for the modelling of populations where subgroups can disband and reform over time, as is the case with many social animals (Oro, 2020). The magnitude of the correlation between changes of one species' abundance and changes in another is a simple way of measuring a trophic system's response to temporal abundance changes, especially when one is solely using observational data. As noted by Berlow et al. (2004), these metrics can sometimes be difficult to interpret, but they do however allow for the inclusion of non–trophic interactions, for example.

\vspace{3mm}

It is desirable to have analytic results characterising the group correlation structure. However, obtaining such results in the general framework considered thus far is likely to be analytically intractable. To do so would involve deriving the (marginal) temporal group correlation
\begin{eqnarray*}
    \rho_{ijt} := \mbox{Corr}(X_{i,t}, X_{j,t}) := \frac{\mbox{Cov}(X_{i,t}, X_{j,t})}{\sqrt{\mbox{Var}(X_{i,t})\mbox{Var}(X_{j,t})}},
\end{eqnarray*} where $\mbox{Corr}(X_{i,t}, X_{j,t})$ is the standard Pearson's correlation between groups $i$ and $j \neq i$ at time $t$. In order to approximate $\rho_{ijt}$, by the Law of Total Expectation, we would have to evaluate \begin{eqnarray*}
        \mathbb{E}[X_{i,t}] &=& \mathbb{E}[\mathbb{E}[X_{i,t} | w, \mbox{$\cal{H}$}_{t-1}]] \\ &=& \mathbb{E}\left[\exp\left(\omega_i^X + \Lambda_{i,t-1}^X+\Psi_{i,t-1}+\Delta_{i,t}\right)\right] \\
        &=& \mathbb{E}\left[\exp \left(\omega_i^X\right)(X_{i,t}+1)^{\lambda_i}\left(\sum_{j \neq i}X_{i,t}+1\right)^{\psi_i}(Y_{t}+1)^{\delta_i}\right].
\end{eqnarray*} This would involve marginalising the product inside this expectation over $w$ and $\cal{H}$$_{t}$. The form of this expectation includes several nonlinearities and approximating it would lead to calculating the expected value of the product of non-independent random variables via sequential Taylor approximations. Even though $X_{it}$ is conditionally Poisson, this does not imply that the marginal distribution of $X_{it}$ is Poisson. In fact it may be a mixture of distributions. This intractability motivated us to approximate $\rho_{ijt}$ under reasonable assumptions. 

\vspace{3mm}

While the general problem seems analytically intractable, it is however possible to derive results which can give insight into the group correlation structure with the addition of some statistically reasonable assumptions. We next outline these assumptions, and show how they allow us to derive an approximation to $\rho_{ijt}$. Following the derivation of an approximation to $\rho_{ijt}$, we will then discuss its interpretation as an indicator of interaction, which is in line with the interaction strength metrics as reviewed in Berlow et al. (2004). A further investigation will have to be carried out in order to explicitly derive the temporally varying (marginal) correlation structure between groups. 

\vspace{3mm}

Modelling the log of abundance using classical autoregressive models, with standard white noise error, is a common method in ecological time series analyses (Ives et al., 2003; O’Hara \& Kotze, 2010; Ovaskainen et al., 2017). So, firstly we assume that
 \begin{equation} \label{modelapprox}
    \begin{aligned} 
        &\ln(Y_{t})| w, \mbox{$\cal{H}$}_{t-1} \sim \mbox{$\cal{N}$}\left( \ln(\mu_{t}^{Y}), \ln(\mu_{t}^{Y}) \right) \\
	    &\ln(X_{i,t})| w, \mbox{$\cal{H}$}_{t-1} \sim \mbox{$\cal{N}$}\left(\ln(\mu_{i,t}^{X}), \ln(\mu_{i,t}^{X}) \right),
    \end{aligned}
\end{equation}
where $\ln(\mu_{t}^{Y})$ and $\ln(\mu_{i,t}^{X})$ are the same as in (\ref{loglink2}) and (\ref{loglink1}). We assume that both $\ln(X_{i,t})$, $\ln(Y_{t})$, $X_{i,t}$ and $Y_{t}$ have finite moments. As we are modelling multiple groups of the same species, we further assume a Heterogeneous Compound Symmetry (HCS) conditional variance-covariance structure (Thall and Vail, 1990; Wolfinger, 1996), similar to when one is analysing repeated measurements of clustered data. That is, for $\sigma_{it}^2 := \mbox{Var}(X_{i,t} | w, \mbox{$\cal{H}$}_{t-1})$, we are assuming that \begin{eqnarray} \label{cov}
    \mbox{Cov}(X_{i,t}, X_{j,t} | w, \mbox{$\cal{H}$}_{t-1}) = \rho_{t} \sigma_{it} \sigma_{jt}  
\end{eqnarray} where $\rho_{t} \in [-1,1]$, i.e. $\rho_{ijt} = \rho_{t}$ for all $i,j \in \{1,...,g\}$, $i \neq j$. Deriving the expectation of non-linear functions of random variables can lead to complicated expressions of covariance and correlations. For $X$ a nonnegative random variable with finite moments, we approximate the marginal expectation of $\ln(X+1)$ using a second order Taylor expansion around zero (Casella and Berger, 2021), which gives  \begin{eqnarray} \label{approxln} \mathbb{E}[\ln(X+1)] \approx \ln(\mathbb{E}(X)+1)+\frac{\mbox{Var}(X)}{2(\mathbb{E}(X) +1)^2}.\end{eqnarray}
As a fourth assumption we assume (\ref{modelapprox}) is weakly stationary. A discrete-time stochastic process $(Z_t)_{[0,T]}$ is said to be weakly stationary if \begin{enumerate}
    \item $\mathbb{E}[Z_t^2] < \infty$ for all $t \geq 0$;
    \item $\mathbb{E}[Z_t] = z \in \mathbb{R}$ for all $t \geq 0$; and
    \item Cov$(Z_t, Z_s)$ = Cov$(Z_{t+r}, Z_{s+r})$ for all $r \geq 0$ and $0 \leq t < s$.
\end{enumerate}
In our framework weak stationarity implies that
\begin{eqnarray*}
    &&\mathbb{E}[\ln(Y_{t}) | w, \mbox{$\cal{H}$}_{t-1}] = \ln(\mu_{t}^{Y}) \equiv \ln(\mu^{Y}) \in \mathbb{R}, \\ 
    &&\mathbb{E}[\ln(X_{i,t}) | w, \mbox{$\cal{H}$}_{t-1}] = \ln(\mu_{i,t}^{X}) \equiv \ln(\mu_i^{X}) \in \mathbb{R}, \ i \in \{1,...,g\}.
\end{eqnarray*} Although many real-world time series are nonstationary, stationarity is a common assumption in ecological time series analyses (Ives, Abbott \& Ziebarth, 2010;  Certain, Barraquand \& Gårdmark, 2018). It is reasonable if one does not want to favour any particular part of the coupling processes or transient dynamics, for example. Weak stationarity then implies that $\rho_{t} \equiv \rho \in [-1,1]$, i.e. we are essentially calculating the net group interactions within a stationary population over the time horizon $[0,T]$. 

\vspace{3mm}

For $\ln(Z) \sim $ $\cal{N}$$(\mu_z, \sigma_z^2)$, the mean and variance of $Z$ are respectively given by: \begin{equation} \label{lognormal}
    \begin{aligned}
        &\mathbb{E}[Z] = \exp\left(\mu_z + \frac{\sigma_z^2}{2}\right),\\ 
&\mbox{Var}(Z) = \exp\left(2\mu_z+\sigma_z^2\right)\left(\exp\left(\sigma_z^2\right)-1\right).
    \end{aligned}
\end{equation} It follows from  (\ref{modelapprox}) and (\ref{lognormal}) that
\begin{eqnarray} \label{condexp1}
    \mathbb{E}\left[X_{i,t} | w, \mbox{$\cal{H}$}_{t-1}\right] = \exp\left(\frac{3\ln\left( \mu_i^X\right)}{2} \right) = \left(\mu_i^X\right)^{3/2}.
\end{eqnarray} Because of our stationarity assumption, the Law of Total Expectation and (\ref{condexp1}) we have that
\begin{eqnarray} \label{condexp2}
   \mathbb{E}[X_{i,t}] = \mathbb{E}\left[\mathbb{E}\left[X_{i,t} |  w,\mbox{$\cal{H}$}_{t-1}\right]\right] = \left(\mu_i^X\right)^{3/2}.
\end{eqnarray} It follows from (\ref{lognormal}), (\ref{condexp2}) and the Law of Total Variance (Klenke, 2013) that \begin{equation} \label{varmu}
\begin{aligned}
    \mbox{Var}(X_{i,t}) &= \mathbb{E}\left[\mbox{Var}(X_{i,t} | w,\mbox{$\cal{H}$}_{t-1})\right] + \mbox{Var}\left(\mathbb{E}\left[X_{i,t} | w,\mbox{$\cal{H}$}_{t-1}\right]\right) \\  &= \mathbb{E}\left[\exp\left(2\ln\left(\mu_i^X\right) + \ln\left(\mu_i^X\right)\right)\left(\exp\left(\ln\left(\mu_i^X\right)\right)-1\right)\right] + \mbox{Var}\left(\left(\mu_i^X\right)^{3/2}\right) \\ 
    &= (\mu_i^X-1)(\mu_i^X)^3 =: V_i^X.
\end{aligned}
\end{equation} In (\ref{varmu}) we use the fact that the variance of a constant is zero. Analogous expressions for the expectation and variance of $Y$ follow similarly. Similarly define $V^Y := (\mu^Y-1)(\mu^Y)^3$. It follows from (\ref{cov}), (\ref{condexp2}) and the Law of Total Covariance (Klenke, 2013) that \begin{equation} \label{covmu}
\begin{aligned}
    \mbox{Cov}(X_{i,t},X_{j,t}) &= \mathbb{E}\left[\mbox{Cov}\left(X_{i,t},X_{j,t} | w,\mbox{$\cal{H}$}_{t-1}\right)\right] \\
    &  + \mbox{Cov}\left(\mathbb{E}\left[X_{i,t} | w,\mbox{$\cal{H}$}_{t-1}\right] \mathbb{E}\left[X_{j,t} | w,\mbox{$\cal{H}$}_{t-1}\right]\right) \\
&= \mathbb{E}\left[\rho \sqrt{V_i^XV_j^X}\right] + \mbox{Cov}\left(\left(\mu_i^X\right)^{3/2}, \left(\mu_j^X\right)^{3/2}\right) \\ 
&= \rho \sqrt{V_i^XV_j^X}. 
\end{aligned}
\end{equation} 

\vspace{3mm}

It follows from (\ref{varmu}), (\ref{covmu}) and Bienaymé's identity (Klenke, 2013) that \begin{eqnarray} \label{bcov}
    \mbox{Var}\left(\sum_{j \neq i}^g X_{j,t}\right) = \sum_{j \neq i}^g V_j^X + \rho\sum_{\substack{k, j \neq i \\ k \neq j}}^g \sqrt{V_j^X V_k^X},
\end{eqnarray} for a fixed $i \in \{1,...,g\}$. By the Law of Total Expectation we have that \begin{eqnarray} \label{lncondexp} \mathbb{E}[\ln (X_{i,t})] = \mathbb{E}\left[\mathbb{E}\left[\ln(X_{i,t})| w, \mbox{$\cal{H}$}_{t-1}\right]\right] \ \forall \ t \in [0,T].\end{eqnarray} Recall that \begin{equation} \label{recall}
\begin{aligned}
    \ln\left(\mu_{it}^X\right) &= \omega^{X}_{i} + \Lambda^{X}_{i,t-1} + \Psi_{i,t-1} + \Delta_{i,t-1} \\
    &= \omega^{X}_{i} + \lambda_i^X\ln\left( X_{i,t-1} +1\right) + \psi_i \ln \left( \sum_{j \neq i}^gX_{j,t-1} +1\right) + \delta_i \ln \left(Y_{t-1} +1\right).
\end{aligned}
\end{equation} The elements of $w\cup\mbox{$\cal{H}$}_{t-1}$ are mutually independent. Therefore because of (\ref{recall}), our assumption of stationarity and the linearity of expectation we have that
\begin{equation} \label{approxmean1}
\begin{aligned}
     \mathbb{E}\left[\mathbb{E}\left[\ln(X_{i,t})| w, \mbox{$\cal{H}$}_{t-1}\right]\right] &= \mathbb{E}\left[\omega^{X}_{i} + \Lambda^{X}_{i,t-1} + \Psi_{i,t-1} + \Delta_{i,t-1}\right] \\ &= \mathbb{E}\left[\omega_i^X\right] + \mathbb{E}\left[\lambda_i^X\right] \mathbb{E}\left[\ln\left( X_{i,t-1} +1\right)\right] \\ &+ \mathbb{E}\left[\psi_i\right]\mathbb{E}\left[\ln \left( \sum_{j \neq i}^gX_{j,t-1} +1\right)\right] \\ &+ \mathbb{E}\left[\delta_i\right] \mathbb{E}\left[ \ln \left(Y_{t-1} +1\right)\right].
    \end{aligned}
\end{equation} By (\ref{lncondexp}) and our stationarity assumption we can equate $\ln(\mu_i^X)$ with the final expression in (\ref{approxmean1}). We already know the random effects, given by (\ref{re}), are Gaussian with their respective means and variances given in (\ref{meanvar}). Using (\ref{bcov}), (\ref{lncondexp}), (\ref{approxmean1}) and our Taylor approximation (\ref{approxln}) we get that \begin{equation*} \label{approxmean2}
\begin{aligned}
    \ln(\mu_i^{X}) &\approx \mu_{\omega} + \mu_{\lambda} \left( \ln\left(\left(\mu_i^X\right)^{3/2}+1\right) + \frac{V_i^X}{2\left(\left(\mu_i^X\right)^{3/2}+1 \right)^2} \right) \\ &+ \mu_{\psi}\ln\left(\sum_{j \neq i}^g\left(\mu_j^X\right)^{3/2}+1\right) + \mu_{\delta} \left( \ln\left(\left(\mu^Y\right)^{3/2}+1\right) + \frac{V^Y}{2\left(\left(\mu^Y\right)^{3/2}+1 \right)^2}\right) \\ &+  \frac{\mu_{\psi}}{2\left(\sum_{j \neq i}^g \left(\mu_j^X\right)^{3/2} + 1 \right)^2} \left(\sum_{j \neq i}^g V_j^X + \rho\sum_{\substack{k, j \neq i \\ k \neq j}}^g \sqrt{V_j^X V_k^X}\right).
    \end{aligned}
\end{equation*} We can then solve for $\rho$ which gives \begin{equation} \label{rhotilde}
\begin{aligned}
\rho \approx \tilde{\rho} &= -\dfrac{A_i}{\mu_{\psi}}-\dfrac{\mu_{\delta}B_i}{\mu_{\psi}}-C_i,
\end{aligned}
\end{equation} for each $i \in \{1,...,g\}$, where \begin{equation*} \label{ABC}
\begin{aligned}
&A_i := -D_i\left(\ln(\mu_i^X) - \mu_{\omega} - \mu_{\lambda}\left(  \ln\left(\left(\mu_i^X\right)^{3/2}+1\right) + \frac{V_i^X}{2\left(\left(\mu_i^X\right)^{3/2}+1 \right)^2} \right)\right),\\
&B_i := D_i \left(\ln\left(\left(\mu^Y\right)^{3/2}+1\right) + \frac{V^Y}{2\left(\left(\mu^Y\right)^{3/2}+1 \right)^2}\right),\\
&C_i := \frac{\sum_{j \neq i}^g V_j^X + \ln\left( \sum_{j \neq i}^g \left(\mu_j^X\right)^{3/2}+1\right) \left(2\left(\sum_{j \neq i}^g \left(\mu_j^X\right)^{3/2} + 1 \right)^2\right)}{\sum_{\substack{k, j \neq i \\ k \neq j}}^g\sqrt{V_j^X V_k^X}}, \\
&D_i := \dfrac{2\left(\sum_{j \neq i}^g \left(\mu_j^X\right)^{3/2} + 1 \right)^2}{\sum_{\substack{k, j \neq i \\ k \neq j}}^g\sqrt{V_j^X V_k^X}}.
\end{aligned}
\end{equation*}

\subsection{Theoretical Interpretation} 
 
We interpret $\tilde\rho$ as the approximate net group correlation. In order that the log-normal approximation is well defined we assume that $\mu_i^X, \mu^Y > 0$ for all $i \in \{1,...,g\}$. In order for both $D_i$ to be well defined we also assume that $\mu_{\psi} \neq 0$. In practice elements of $M$ will not be exactly $0$ and so this is reasonable to assume.  In order for both $C_i$ and $D_i$ to be well defined, we assume that, for all $i \in \{1,...,g\}$, \[V_i^X = (\mu_i^X-1)(\mu_i^X)^3>0  \iff \mu_i^X>1.\]

As $\tilde\rho$ is not indexed by $i$ or $t$, we thus have $g$ approximations to the net interactions between groups over $[0,T]$. Note that (\ref{rhotilde}) has a dependence on $i$ on the right hand side.  As $\tilde{\rho}$ has no such dependence, we effectively have $g$ different approximations for the same quantity. If our underlying assumptions are correct, these should be consistent and not vary significantly.  This simple observation gives us a way of testing our assumptions and, where these appear to be valid, we could potentially combine the independent approximations to obtain an improved estimate of the net group correlation. We also note that since this approximation is based on a second order Taylor expansion around zero, and other simplifying assumptions, $\tilde\rho$ may not be contained within $[-1,1]$ for certain parameter and mean values. However, should the assumptions hold this approximation would be an informative measure of net group interactions. If this approximation resulted in $\tilde\rho$ being slightly outside $[-1,1]$ one could for example take the approximate net group correlation as being given by \begin{eqnarray*}
    \tilde\rho^* := \left\{ \begin{array}{l}
         1, \mbox{ if $\tilde\rho>1$} \\
         \tilde\rho, \mbox{ if $\tilde\rho \in [-1,1]$} \\
          -1, \mbox{ if $\tilde\rho<-1$}.
    \end{array} \right.
\end{eqnarray*}

As our approximation assumes weak-sense stationarity, it would not be appropriate for the wolf-elk application discussed above. As seen in Fig. \ref{wolfelkplot} the time series used in our case study are highly nonstationary. In this application we also include a pack formation-splitting process, which for certain $i$ and $t$ results in supp$(X_{it})=0$, i.e. for certain groups and at certain times $X_{it}$ is modelled by a degenerate Poisson process. Instead, for illustrative purposes we will theoretically interpret $\tilde{\rho}$ in the context of a general predator-prey system, i.e. $Y$ are prey and $X_1,...,X_g$ are predator groups. 

\vspace{3mm}

The signs of the elements of $M:= \{\mu_{\omega}, \mu_{\lambda}, \mu_{\psi}, \mu_{\delta}\}$ can be either positive or negative. Therefore there are $2^4 = 16$ possible positive-negative parameter combinations. The magnitude of these values can also result in additional scenarios to discuss. So, for brevity and illustrative purposes, we will look at four ecologically meaningful scenarios. We will discuss how the intraspecific and auxiliary coupling parameters ($\mu_{\psi}$ and $\mu_{\delta}$) play a role in determining the sign of $\tilde \rho$. In each scenario we assume that $\mu_{\omega}>0$ (positive population-level growth rate) and $\mu_{\lambda} \geq 1$ (strong population density dependence). 

\vspace{3mm}

Clearly $0<x<1$ implies that $\ln(x)<0$. It is not difficult to see that, by monotonicity of $\ln$, \[x>1 \implies \ln\left(x^{3/2}+1\right)>\ln\left(x+1\right)>\ln(x)>0.\] For $\mu_{\lambda} \geq 1$, it is then clear to see that  \begin{eqnarray} \label{log}
       \ln\left(x\right)- \mu_{\lambda} \ln\left(x^{\frac{3}{2}}+1\right)<0, \ \forall \ x>0. 
\end{eqnarray} It then follows that \[\mu^X_i, \mu^Y > 0 \implies D_i, A_i, B_i, C_i>0, \ \forall \ i \in \{1,...,g\}.\]

\begin{figure}[!h]
\centering
\includegraphics[trim = 195 70 260 70, clip, width = 10.5cm]{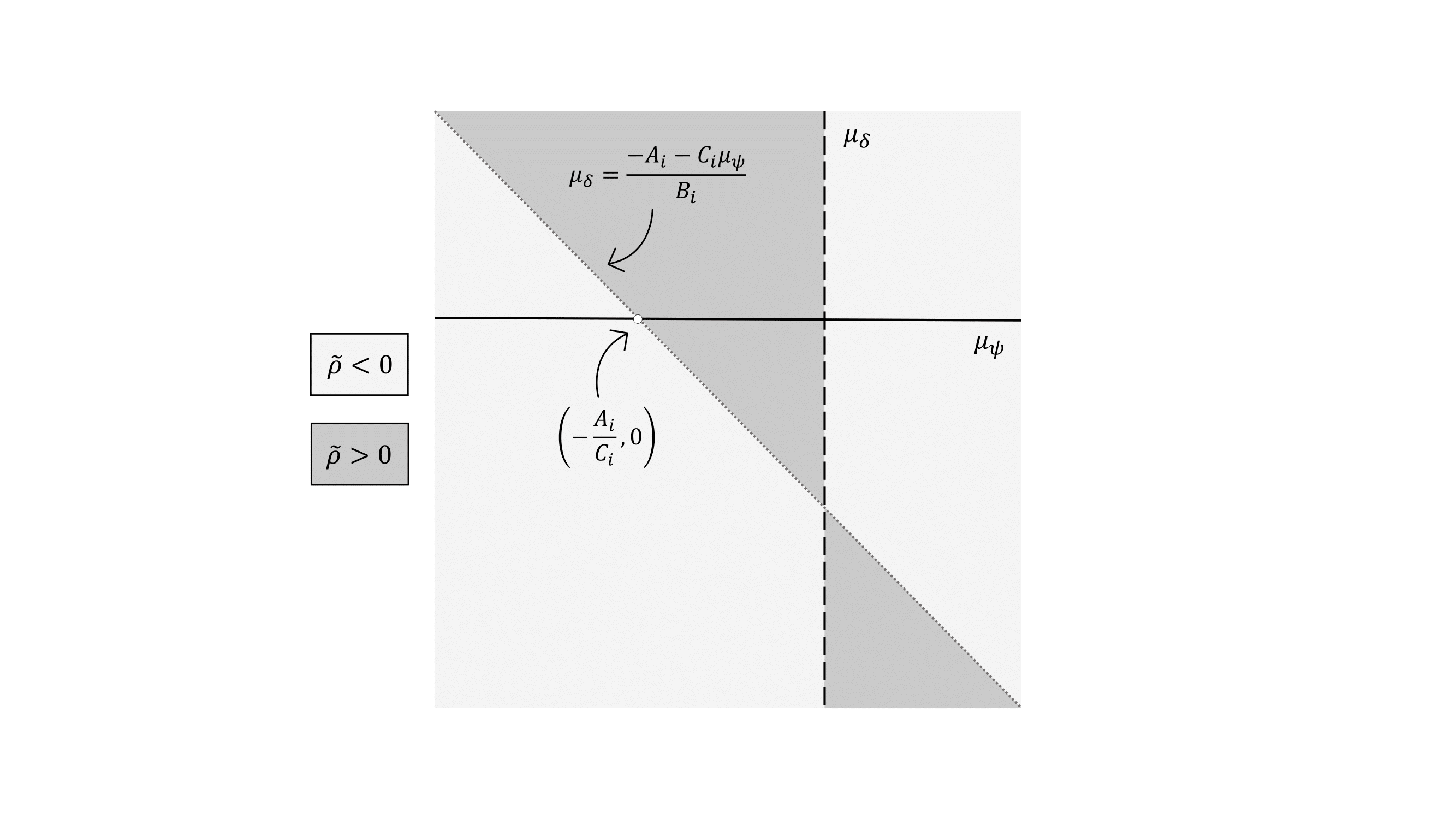}
\caption{Illustration of the $\mu_{\psi}$-$\mu_{\delta}$ plane for some fixed $i \in \{1,...,g\}$, treating all parameters except $\mu_{\psi}$ and $\mu_{\delta}$ as known and fixed. Dark (light) grey shaded areas indicate where $\tilde \rho >0$ ($\tilde \rho <0$). The grey dotted line represents the line $-A_i-\mu_{\delta}B_i-\mu_{\psi}C_i=0$. The black dashed line indicates that $\mu_{\psi}\neq0$ (by assumption). The white circle is the point $(\mu_{\psi}, \mu_{\delta}) = (-A_i/C_i,0)$.}
\label{abcfig}
\end{figure}

From Fig. \ref{abcfig} we can see that if $\mu_{\psi}>0$, for each $i \in \{1,...,g\}$, \begin{eqnarray*}
    &&\tilde \rho >0 \iff A_i+\mu_{\delta}B_i+\mu_{\psi}C_i <0 \\
    &&\tilde \rho <0 \iff A_i+\mu_{\delta}B_i+\mu_{\psi}C_i >0,
\end{eqnarray*} and if $\mu_{\psi}<0$, for each $i \in \{1,...,g\}$,  \begin{eqnarray*}
    &&\tilde \rho >0 \iff A_i+\mu_{\delta}B_i+\mu_{\psi}C_i >0 \\
    &&\tilde \rho <0 \iff A_i+\mu_{\delta}B_i+\mu_{\psi}C_i <0.
\end{eqnarray*}

\textbf{(P1)} First, suppose $\mu_{\delta}, \mu_{\psi} > 0$. The sign of $\mu_{\delta}$ suggests that the effect of increasing group abundance has a net benefit to the mean processes of each group. The prey coupling and benefit of increasing group abundance will be stronger as both of these parameters become large in magnitude. This may arise, for example, in a scenario where the overall predator population is sufficiently large, inter-group responses to increases in predator abundance is strong and the predator is a specialist, i.e. monitors prey availability and is sensitive to abundance changes in their primary prey (Araujo \& Moura, 2022). As $\mu_{\psi}, \mu_{\delta}>0$ and (\ref{log}) holds we have that $\tilde{\rho} <0$ irrespective of the magnitudes of $\mu_{\delta}$ and $\mu_{\psi}$. The sign of $\mu_{\psi}$ suggests that there is a positive coupling between predator groups and their prey. Depending on the values of the stationary means of $X_i$ and $Y$, the magnitude of $\tilde\rho$ will vary. Therefore these group responses may still result in either weak ($\tilde{\rho} \approx 0$) or strong ($\tilde{\rho} \ll 0$) negative interactions. There also could be abundant resources, i.e. multiple prey types, for each group. This in turn can allow for a strong numerical response (Berardo et al., 2020).

\vspace{3mm}

\textbf{(P2)} Assume $\mu_{\psi} < 0$. We have two cases to consider for the sign of $\mu_{\delta}$.
\begin{enumerate}    
    \item Suppose $\mu_{\delta} > 0$. As in \textbf{(P1)}, this means there is a positive coupling between predator population and their prey. With $\mu_{\psi}<0$, the sum of the expression involving $A_i$ and the remaining terms involving $B_i$ and $C_i$ can result in either positive or negative $\tilde\rho$ (see Fig. \ref{abcfig}). For example, if the prey stationary mean is sufficiently large, a situation may arise where the positivity of the first term dominates the negativity of the sum of the other terms, resulting in $\tilde\rho>0$. This may arise, for example, in a scenario where there is specialist predation and strong inter-group responses to changes in predator abundance (potentially due to strong competition). This can still result in positive interactions among groups, with the prey-coupling potentially compensating for the negative effects of inter-group changes in abundance. This could be explained, for example, by hunting cooperation among certain groups (Alves and Hilker, 2017). On the other hand, if there are a large number of predator groups, $-C_i<0$ may be negative enough to result in $\tilde\rho<0$. This may arise in a situation where there are negative interactions at higher group densities, where strong prey-coupling compensates for the large negative effects of overcrowding, for example. This has been shown theoretically by Jin \& Wang (2017) for a particular spatial model of prey-taxis (where predators have a tendency to move toward regions of highest prey density).

    \vspace{3mm}
    
    \item Suppose $\mu_{\delta} < 0$. This means that as prey abundance increases this will have negative net effects on predator abundance. This may arise in a scenario where there is predator satiation (Lehtonen \& Jaatinen, 2016). This is an inverse density-dependent process, where a prey population inundates their predators with more food than they can consume at a certain time. This is followed by little to no prey reproduction, thus causing large decreases in the numerical responses of the predator. This can occur via synchronised birth events, herding behaviour or seed masting for example (Tsvuura et al., 2011; Diekmann \& Planqué, 2020; Menezes, Rangel \& Moura, 2022). In this context we could get that $\mu^Y$ is large in magnitude and could result in $\tilde\rho<0$. When resources become scarce following rapid decreases in prey, the negative group interactions would be induced in response to resource scarcity. On the other hand, if $g$ and $\mu^Y$ were small, and $\mu_{\psi}$ was large enough, it could happen that the expression involving $B_i>0$ may be large enough to result in $\tilde\rho>0$. This could arise in a scenario where the predator is a generalist, i.e. those that consume numerous species while having little dependence on specific prey types (Araujo \& Moura, 2022), and the prey population being modelled decreases due to environmental disturbances (such as anthropogenic pressure or anomalous temperature changes). This would result in positive interactions between groups due to hunting cooperation, in order to switch to alternative prey types. 
\end{enumerate} \textbf{(P3)} Lastly, suppose $\mu_{\delta} \approx 0$ and $\mu_{\psi}>0$ is sufficiently large so that $B_i \approx 0$. As mentioned above, the mean parameters in $M$ are not exactly equal to zero, but here we assume $\mu_{\psi}$ is sufficiently large so that the expressions involving $B_i$ adds little contribution to $\tilde\rho$. This implies that there is weak coupling between predator groups and their prey. It also implies that the effect of increasing group stationary means increases the mean process of each group. This may arise, for example, in a scenario where predation is generalist. These values also imply that inter-group responses to increases in predator abundance is beneficial, which could have multiple interpretations. If there is weak dependence on changes in both the prey population and group abundances, then this could signify that there is prey switching occurring in this system (van Leeuwen et al., 2013; Piltz, Porter \& Maini, 2014). Having abundant resources means that direct competition may be weak, for example. All the while, the effects of predation could remain sufficiently low for the prey to have minimal impact. There could also be a low number of groups, with small spatial overlapping of ranges/territories, and so scarce aggressive interactions. This could consequently lead to competitive effects remaining weak overall and so interactions are deemed beneficial for each group.

\vspace{3mm}

We would like to point out that the above interpretations are not exhaustive. Other scenarios can be extrapolated from these parameter values taking various signs and magnitudes. However, based on theoretical and observed phenomena in ecology, we have discussed four plausible scenarios that could occur. In each of these scenarios the sign of $\tilde\rho$ can vary, with its magnitude also changing in response to changes in both the magnitudes of elements of $M$, given in (\ref{meanvar}), and the magnitude of the stationary means of $X_i$ and $Y$. The strength of group interactions can change depending on the trade-off between processes like resource competition and predation, for example, the balance of which is highly nonlinear (Chesson \& Kuang, 2008). This dichotomy is not so clear, as we can see in the above scenarios. The sign of $\tilde\rho$ is not a direct measure of interaction in the sense of there being strong/weak predation or competition in the system. Incidental events such as indirect mutualism, temperature changes, drought, hunting pressure and other ecological and environmental phenomena, that are not explicitly observed, may also affect the sign of $\tilde\rho$. 

\vspace{3mm}

At certain instances demographic and ecological processes, such as those mentioned in the above scenarios, may either promote or reduce a groups growth, due to events such as discounting (Hauert et al., 2006) or predator interference (Cosner et al., 1999), for example. From an evolutionary game theory perspective, one can view heterogeneous populations as being composed of cooperators (those that make maximal investment to fitness and survival) and defectors (those that do not contribute to fitness and survival). It has been theoretically shown that the ratio of gains to losses, with respect to cooperative and defective interactions, must exceed a threshold value in order for cooperation to evolve. In fact, it has been shown that this ratio must exceed the mean degree of nearest neighbors when one looks at an evolutionary graph of a population, where individuals are nodes and edges are interactions (Lieberman, Hauert \& Nowak, 2005). Thus network reciprocity may come into play in order to maintain cooperation on a population level. That is the clustering of cooperators within a population can protect such groups against defectors. In the above illustrative scenarios the varying interaction strengths, as approximated by $\tilde\rho$, must signify that temporal interactions remain positive on a population level in order for cooperation to exist and for groups to remain intact. But simultaneously these interactions must not be too costly that one group may dominate over the rest and hinder population level fitness.

\section{Summary and Conclusion}
In this paper we proposed a stochastic framework for understanding the dynamics of animal groups within a population in order to infer group interactions/associations, while also accounting for the influence of an auxiliary population. Our general framework does not rely on the assumption that the logarithms of abundance counts are Gaussian and the underlying time series is stationary, which are common assumptions made when modelling species dynamics (Ives et al., 2003; Zhou, Fujiwara \& Grant, 2016). We also model both the observation and latent processes, allowing for the accommodation of overdispersion. We fitted this model to a real-life data-set of wolf-elk time series over 20 years and interpreted parameter values in this context. Our results suggest that wolf predation is not the primary cause of decreases in elk abundance, inter-pack dynamics are largely heterogenous, the grey wolf population in YNP is a generalist predator and inter-pack associations have an overall weak effect of regulating population abundance. Through simulation studies we showed how unbiased and robust our model is when estimating parameters in a variety of contexts. We finally derived an approximation to the inter-group correlation under typical statistical and ecological assumptions, while also discussing its theoretical interpretation in some illustrative predator-prey scenarios. This approximation can capture a diverse range of predator-prey relationships that one can frequently observe in nature such as hunting cooperation and predation satiation, as already discussed. For example, when the net effect of changes in inter-group abundance is negative ($\mu_{\psi}<0$) we stated a sufficient condition based on the ratio of $\mu_{\delta}$ and $\mu_{\psi}$ in terms of the stationary means of $Y$ and $X_i$ that ensures that $\tilde\rho$ is either positive or negative. As mentioned above, the approximation $\tilde\rho$ also serves as a verifying condition for our assumptions to hold and is useful to check if the system one is studying exhibits such features.

\vspace{3mm}

We should point out that there are many opportunities for extending our framework. If one was interested in short term predictions one could use observational random effects in order to account for potential outliers (Oliveira \& Moral, 2021). If one was interested not only in the temporal aspect of group interactions, but also their spatial extent, then one could incorporate a temporally varying network structure into our model. This could be done using, for example, multilayer networks (Hutchinson et al., 2018) or graphical models (Momal, Robin \& Ambroise, 2021). It is also worth noting that if one had higher resolution data they could quantify group interactions via convergent cross mapping, as carried out by Barraquand et al. (2021) for inferring predator-prey cycles for example. One disadvantage of this is the assumption of log-Gaussianity for the abundance random variables. This methodology would have to be improved upon when modelling (potentially low) count time series like in this study.

\section{Acknowledgements}
We would like to thank Ellen Brandell for the conversations had before writing this manuscript. We would also like to acknowledge the ecologists and field researchers who collected the wolf-elk counts in YNP. BMC is supported by an IRC Scholarship $[\mbox{GOIPG}/2020/939]$.

\section{Data Accessibility Statement}
Reproducible R code can be found at: \url{https://github.com/blakeacorrigan/group_interactions.git}. The dataset of Brandell et al. (2021) can be found at: \url{https://datadryad.org/stash/dataset/doi:10.5061%2Fdryad.vq83bk3qd}.

\section{Disclosure}
The authors declare that they have no conflict of interest.

\newpage

\section{Appendix}

\renewcommand{\thefigure}{A1}

In this appendix we present the plots related to our simulation study and case study. Figures A1 - A5 are boxplots of bias and relative bias for our parameter values over all 100 simulations for each S1 - S5, respectively. Bivariate plots for one of our simulation studies (simulation 20) are also given (Figures A6 - A8). These show the estimated posterior density for each parameter and demonstrate when (if present) there was confounding between variables within the MCMC runs. The final plot, A9, is the posterior-prior overlap (PPO) for estimated parameters in the case study, which was used to conclude if certain parameters were weakly identifiable. See the relevant sections in the manuscript for a brief discussion based on these plots. 

\vfill

\begin{figure}[!h]
\centering
\includegraphics[clip, width = 11.8cm]{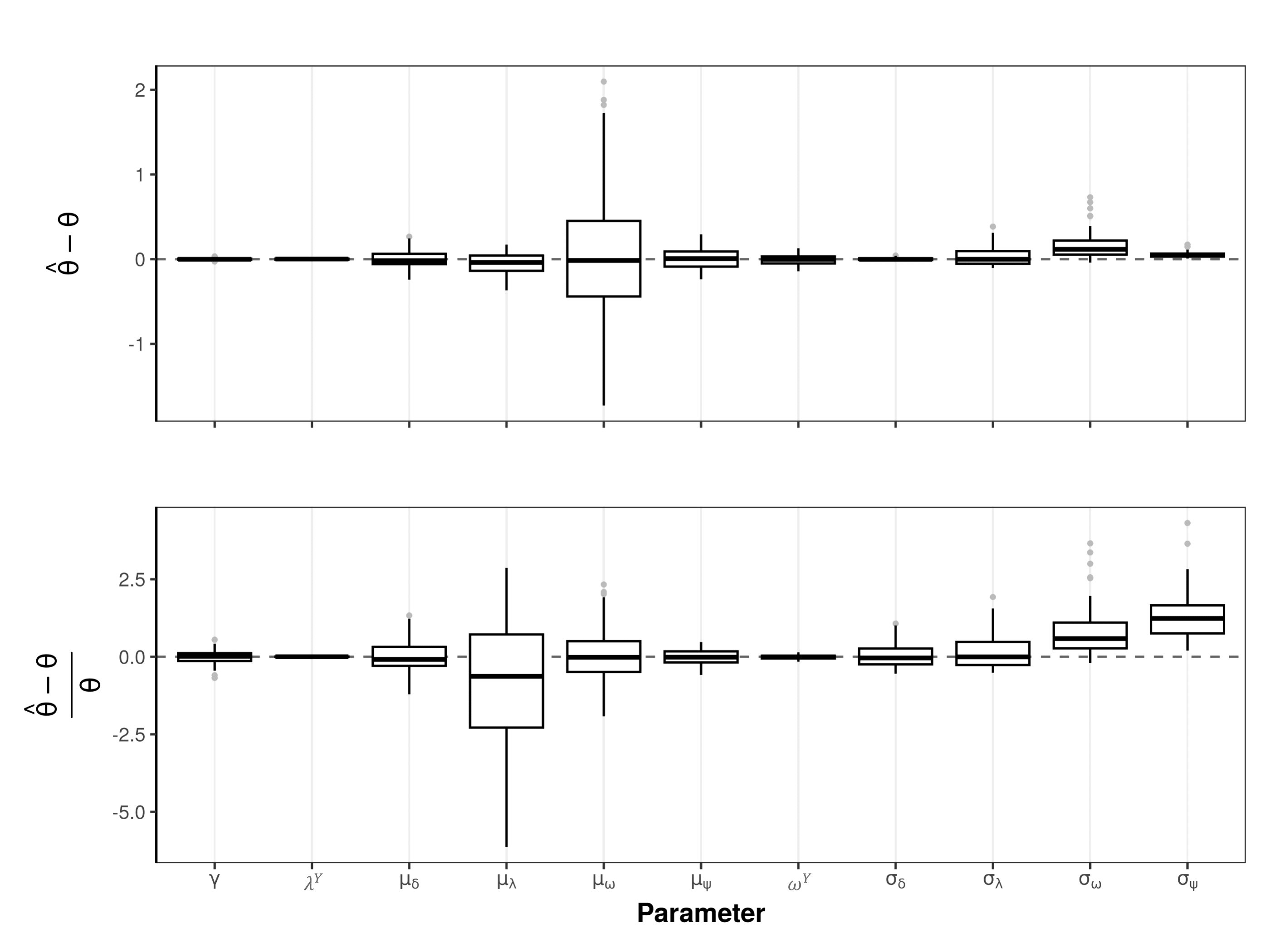}
\caption{Boxplots of bias and relative bias for $\theta \in M \cup \Sigma \cup \{\gamma, \lambda^{Y}, \omega^Y\}$ for S1.}
\label{small_box}
\end{figure}

\vfill

\renewcommand{\thefigure}{A2}

\begin{figure}[!h]
\centering
\includegraphics[clip, width = 11.8cm]{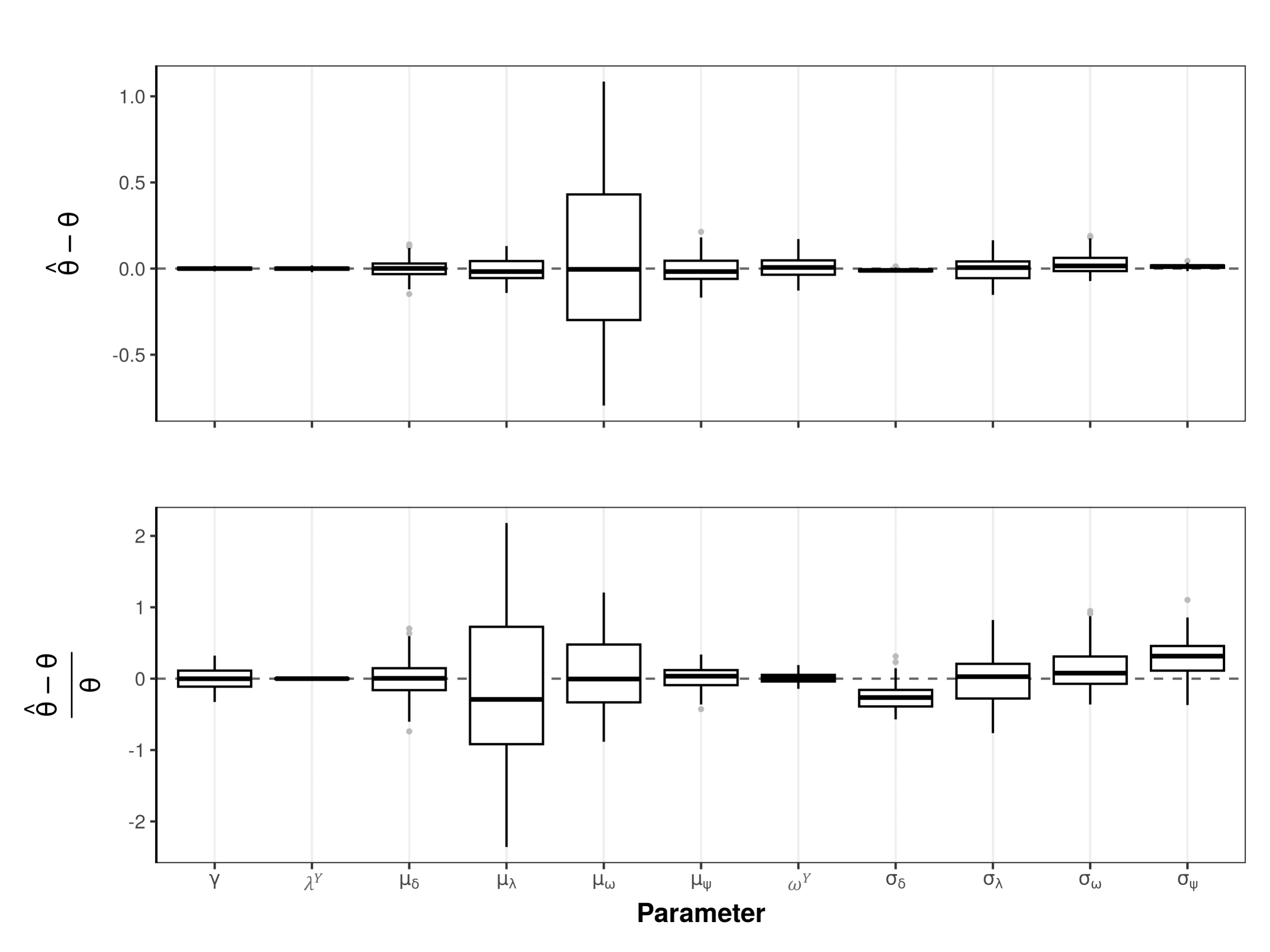}
\caption{Boxplots of bias and relative bias for $\theta \in M \cup \Sigma \cup \{\gamma, \lambda^{Y}, \omega^Y\}$ for S2.}
\label{mod_box}
\end{figure}

\renewcommand{\thefigure}{A3}

\begin{figure}[!h]
\centering
\includegraphics[clip, width = 11.8cm]{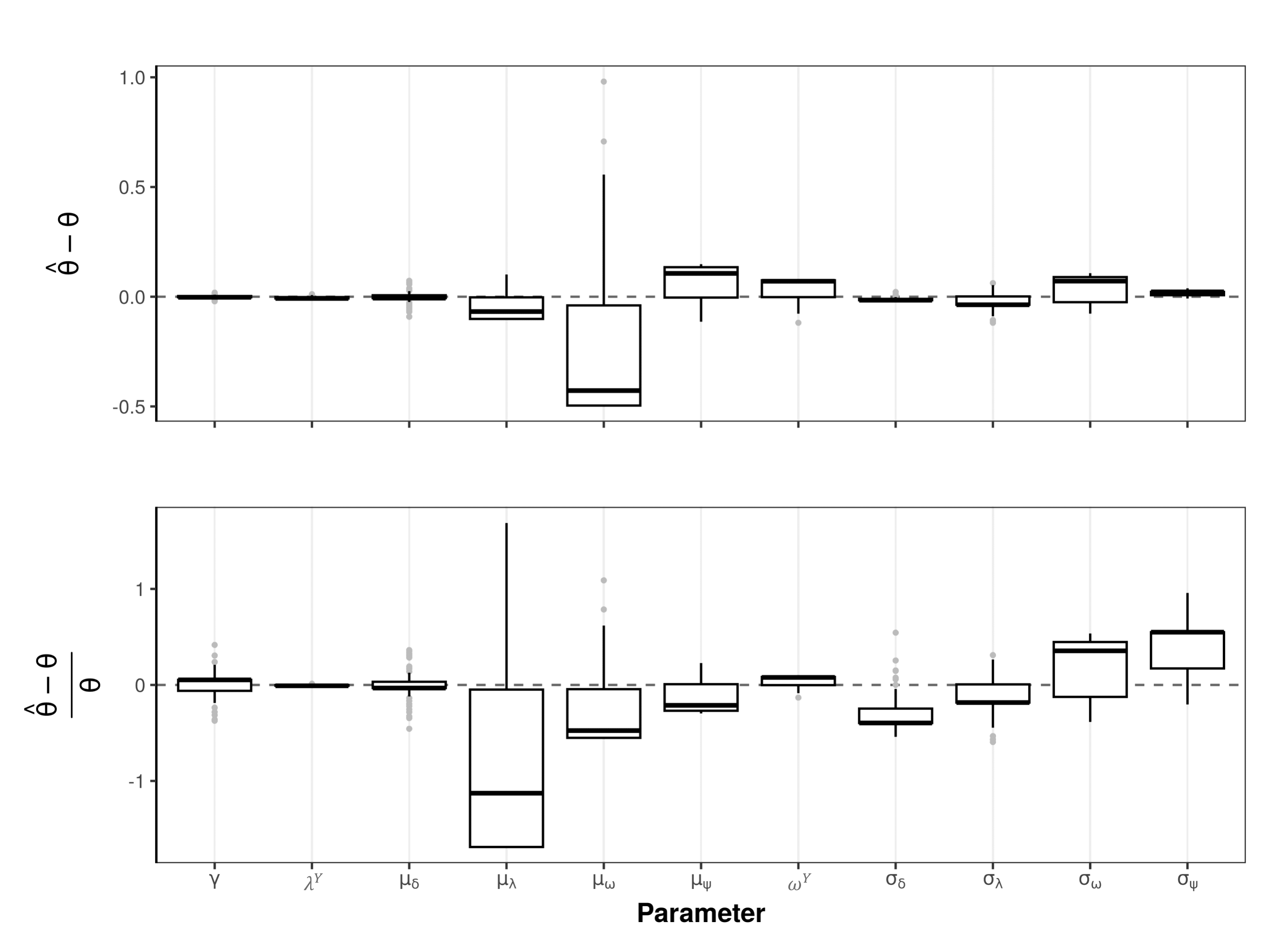}
\caption{Boxplots of bias and relative bias for $\theta \in M \cup \Sigma \cup \{\gamma, \lambda^{Y}, \omega^Y\}$ for S3.}
\label{large_box}
\end{figure}

\renewcommand{\thefigure}{A4}

\begin{figure}[!h]
\centering
\includegraphics[clip, width = 11.8cm]{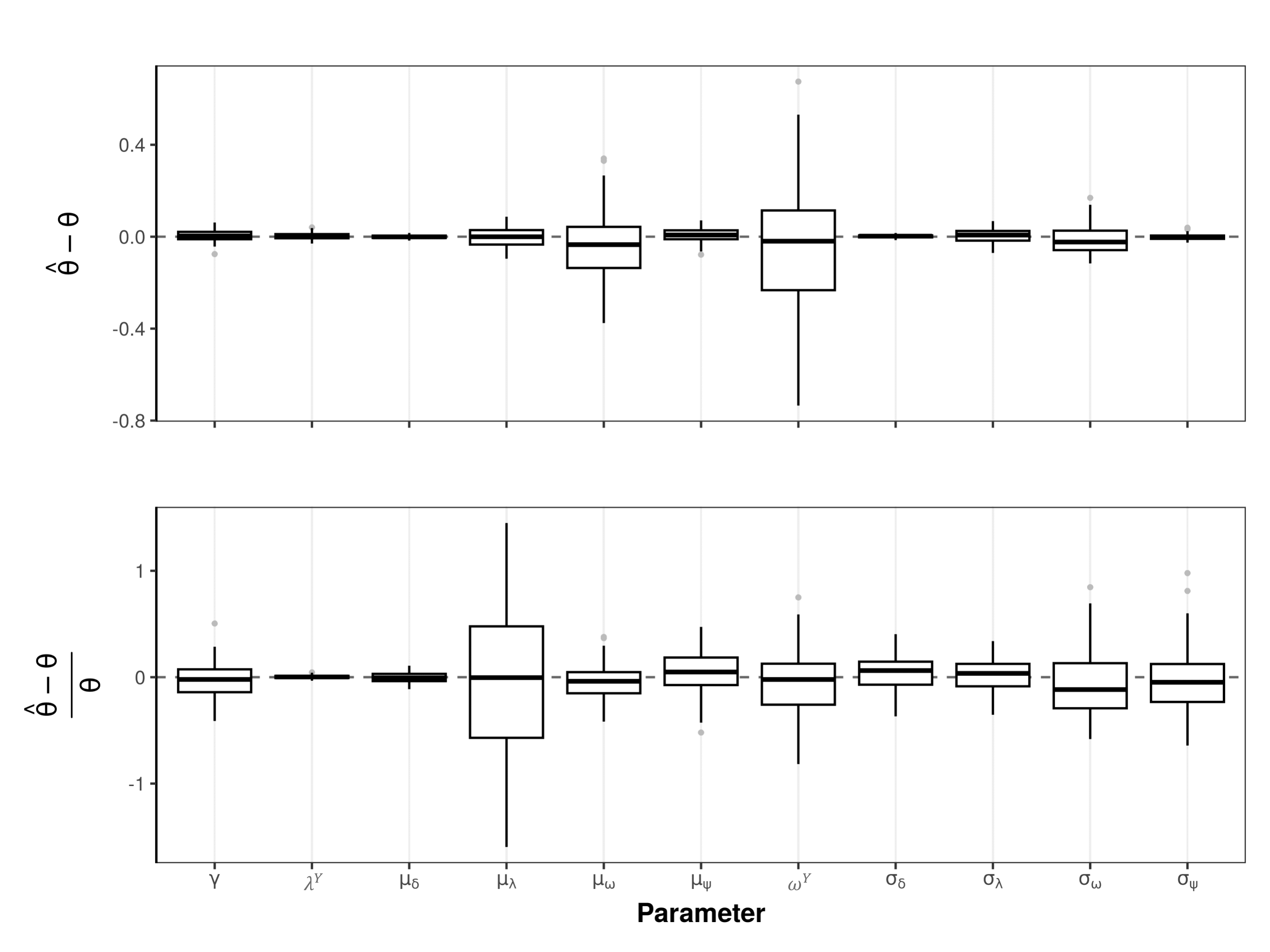}
\caption{Boxplots of bias and relative bias for $\theta \in M \cup \Sigma \cup \{\gamma, \lambda^{Y}, \omega^Y\}$ for S4.}
\label{complow}
\end{figure}

\renewcommand{\thefigure}{A5}

\begin{figure}[!h]
\centering
\includegraphics[clip, width = 11.8cm]{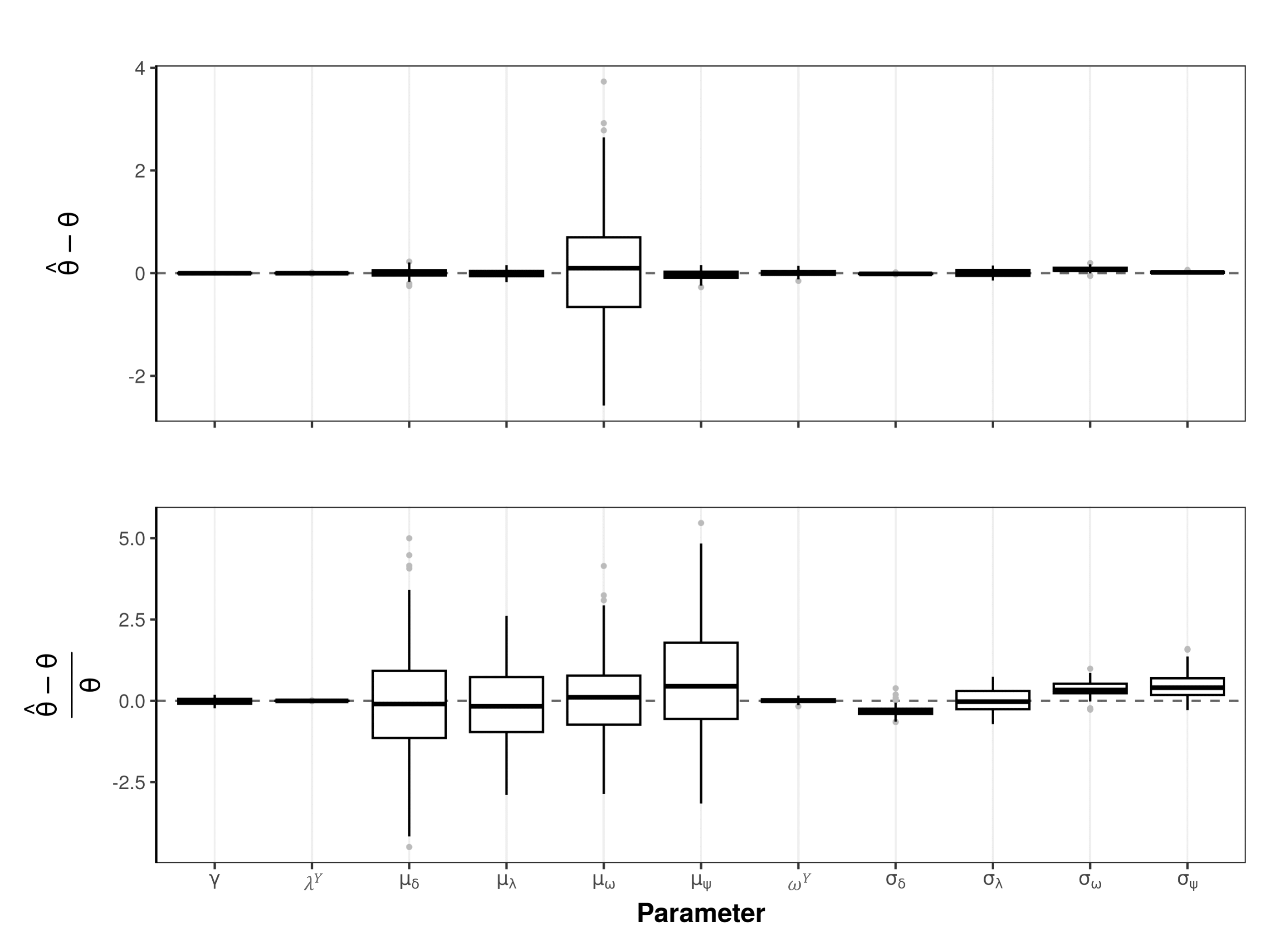}
\caption{Boxplots of bias and relative bias for $\theta \in M \cup \Sigma \cup \{\gamma, \lambda^{Y}, \omega^Y\}$ for S5.}
\label{comphigh}
\end{figure}

\renewcommand{\thefigure}{A6}

\begin{figure}[!h]
\centering
\includegraphics[clip, width = 11.8cm]{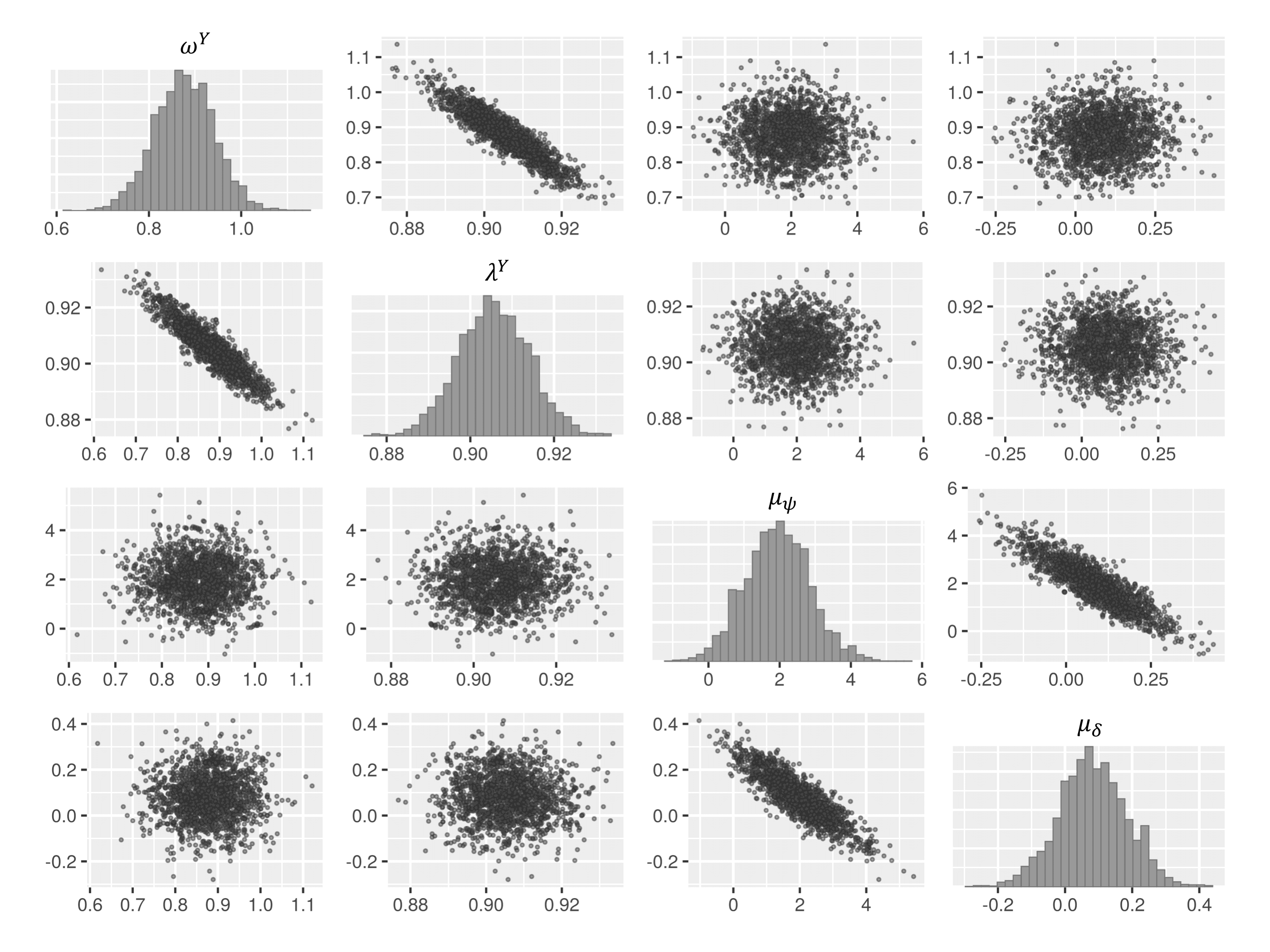}
\caption{Bivariate plots for $\{\omega^Y, \lambda^Y, \mu_{\psi}, \mu_{\delta}\}$ from simulation No. 20 for S1.}
\label{biv1}
\end{figure}

\renewcommand{\thefigure}{A7}

\begin{figure}[!h]
\centering
\includegraphics[clip, width = 11.8cm]{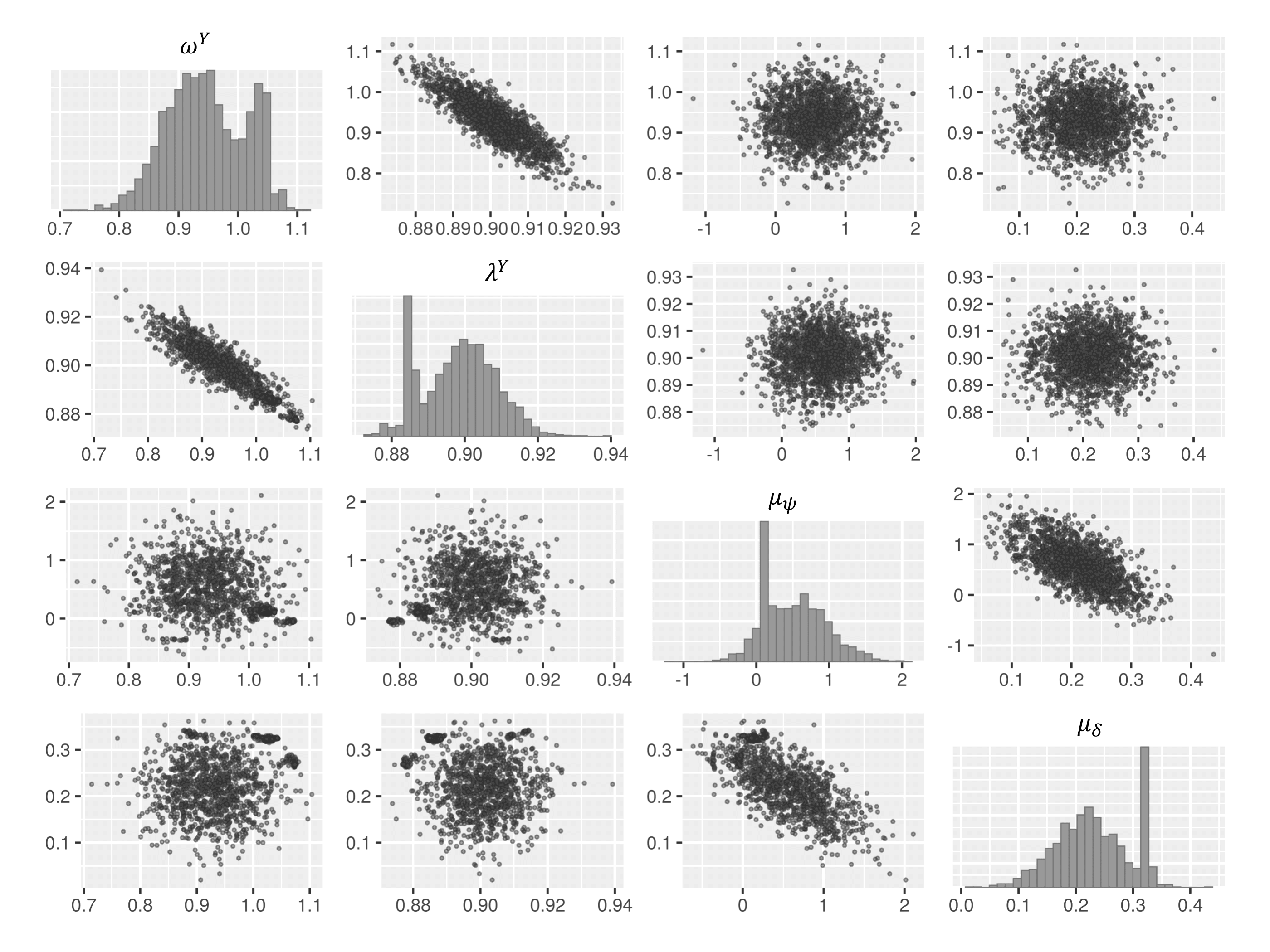}
\caption{Bivariate plots for $\{\omega^Y, \lambda^Y, \mu_{\psi}, \mu_{\delta}\}$ from simulation No. 20 for S2.}
\label{biv2}
\end{figure}

\renewcommand{\thefigure}{A8}

\begin{figure}[!h]
\centering
\includegraphics[clip, width = 11.8cm]{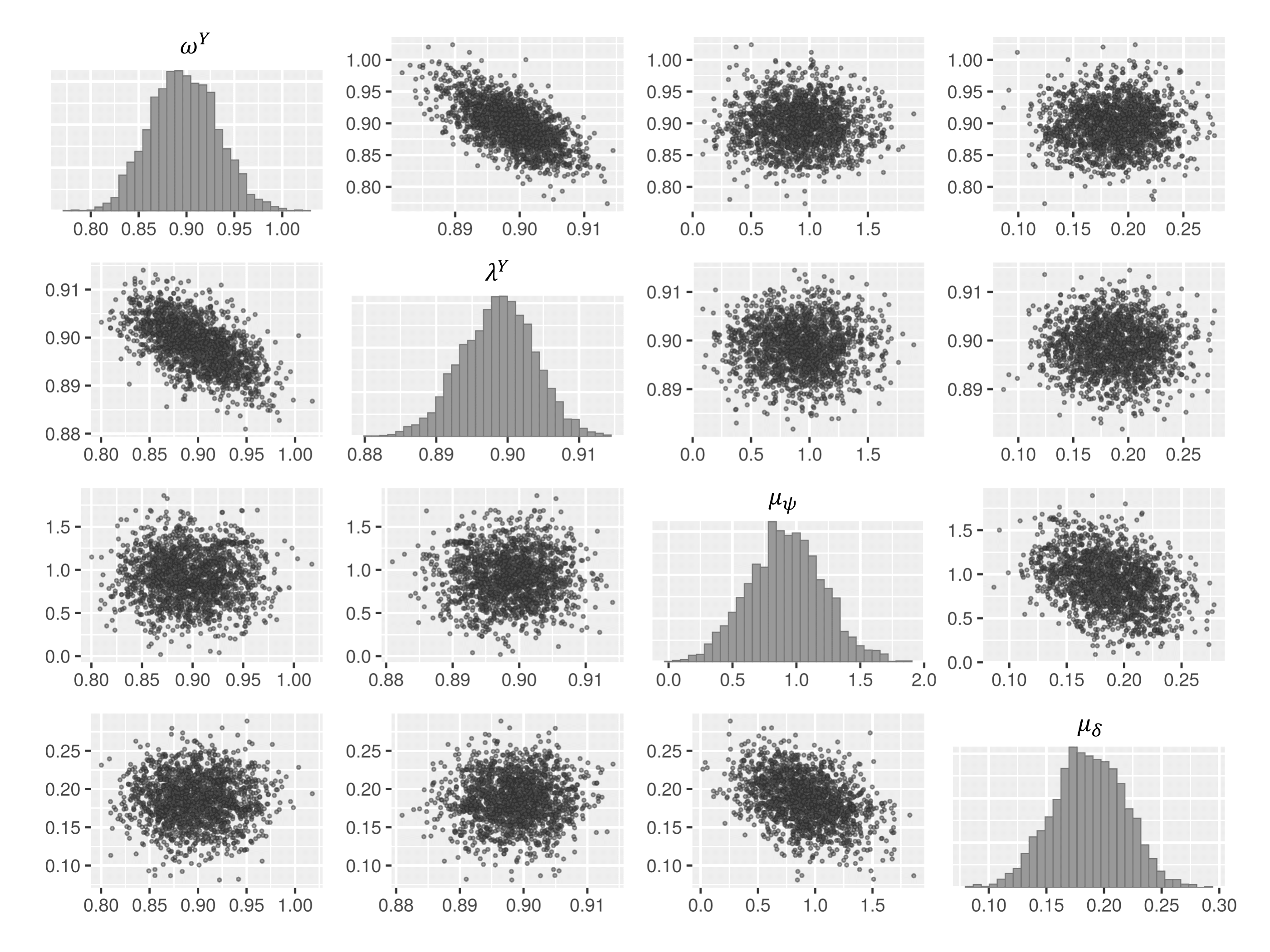}
\caption{Bivariate plots for $\{\omega^Y, \lambda^Y, \mu_{\psi}, \mu_{\delta}\}$ from simulation No. 20 for S3.}
\label{biv3}
\end{figure}

\renewcommand{\thefigure}{A9}

\begin{figure}[!h]
\centering
\includegraphics[clip, width = 11.8cm]{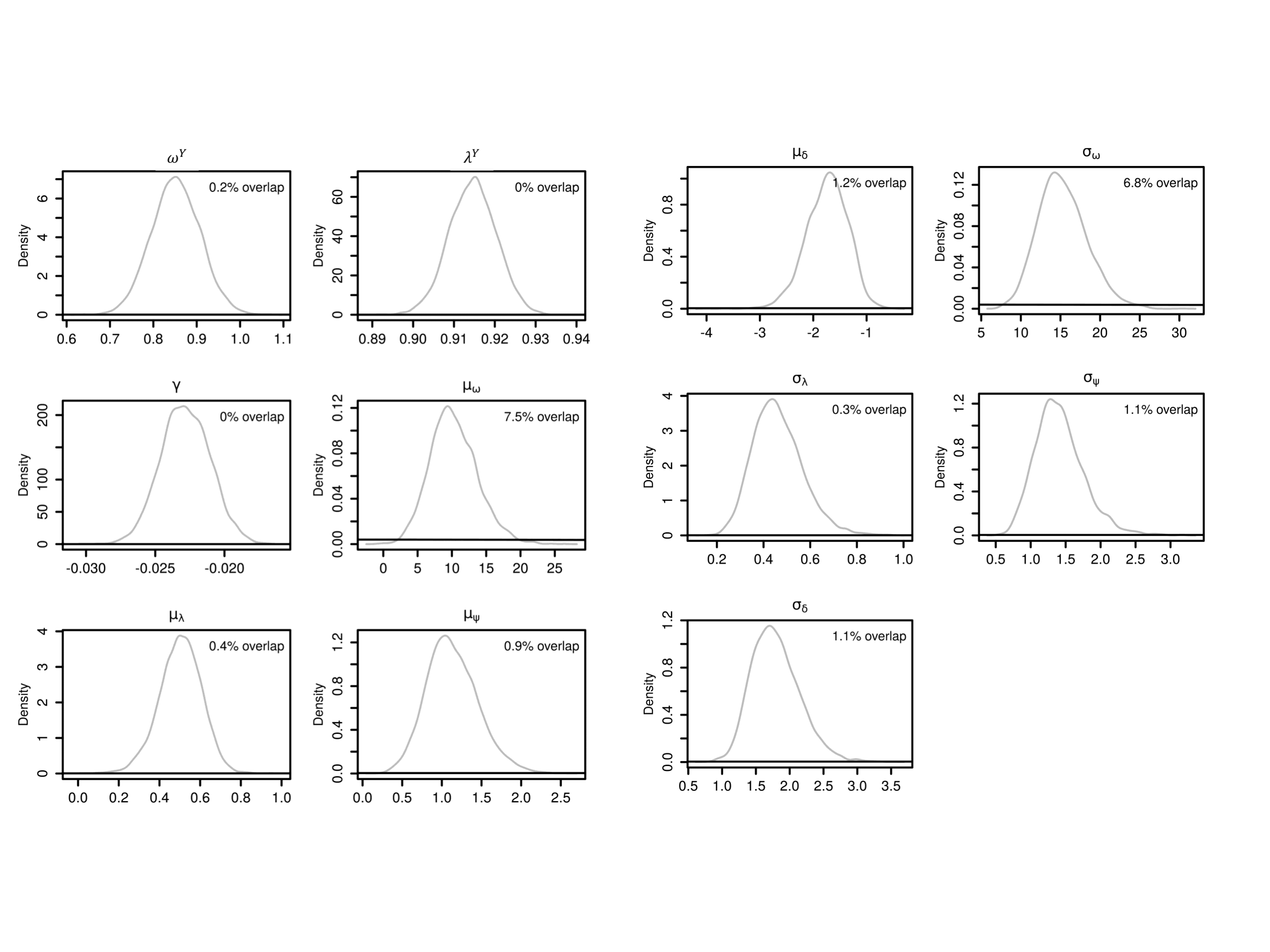}
\caption{Prior- posterior overlap for fixed effects and hyperparameters. The black line is the prior distribution and grey line is estimated posterior density. We set $\theta_{\mu} \sim \mbox{$\cal{N}$}(0,100^2)$ for  $\theta_{\mu} \in \mbox{M} \cup \left\{\omega^Y, \lambda^Y, \gamma \right\}$ and $\theta_{\sigma} \sim \mbox{$\cal{N}$}^+(0,100^2)$, for  $\theta_{\sigma} \in \Sigma$.}
\label{identplot}
\end{figure}

\end{document}